\documentclass[twocolumn, twocolappendix]{aastex631}

\usepackage{amsmath}	
\usepackage{multirow}

\begin{document}

\title{Modeling the Spectral Energy Distribution of Active Galactic Nuclei: Implications for Cosmological Simulations of Galaxy Formation}

\correspondingauthor{Qi Guo}
\email{qguo@bnu.edu.cn}

\author[0009-0000-3702-1954]{Tong Su}
\affiliation{Key Laboratory for Computational Astrophysics, National Astronomical Observatories, Chinese Academy of Sciences, Beijing 100012, China}
\affiliation{School of Astronomy and Space Sciences, University of Chinese Academy of Sciences, 19A Yuquan Road, Beijing 100049, China}
\affiliation{Institute for Frontiers in Astronomy and Astrophysics, Beijing Normal University, Beijing 102206, People’s Republic of China}
\affiliation{School of Physics and Astronomy, Beijing Normal University, Beijing 100875, People’s Republic of China}

\author[0000-0002-7972-3310]{Qi Guo}
\affiliation{Institute for Frontiers in Astronomy and Astrophysics, Beijing Normal University, Beijing 102206, People’s Republic of China}
\affiliation{School of Physics and Astronomy, Beijing Normal University, Beijing 100875, People’s Republic of China}
\affiliation{Key Laboratory for Computational Astrophysics, National Astronomical Observatories, Chinese Academy of Sciences, Beijing 100012, China}
\affiliation{School of Astronomy and Space Sciences, University of Chinese Academy of Sciences, 19A Yuquan Road, Beijing 100049, China}

\author{Erlin Qiao}
\affiliation{Key Laboratory of Space Astronomy and Technology, National Astronomical Observatories, Chinese Academy of Sciences, Beijing 100101, China}
\affiliation{School of Astronomy and Space Sciences, University of Chinese Academy of Sciences, 19A Yuquan Road, Beijing 100049, China}

\author[0009-0000-0690-5562]{Wenxiang Pei}
\affiliation{Shanghai Key Lab for Astrophysics, Shanghai Normal University, Shanghai 200234, People’s Republic of China}

\author[0000-0001-6947-5846]{Luis C. Ho}
\affil{Kavli Institute for Astronomy and Astrophysics, Peking University, Beijing 100871, China}
\affil{Department of Astronomy, School of Physics, Peking University, Beijing 100871, China}

\author[0000-0001-9016-5332]{Cedric G. Lacey}
\affiliation{Institute for Computational Cosmology, Department of Physics, University of Durham, South Road, Durham, DH1 3LE, UK}




\defcitealias{Henriques15}{H15}

\begin{abstract}

Modeling the spectral energy distribution (SED) of active galactic nuclei (AGN) plays a very important role in constraining modern cosmological simulations of galaxy formation. Here, we utilize an physically-motivated supermassive black hole (SMBH) accretion disk model to compute the accretion flow structure and AGN SED across a wide range of black hole mass ($M_\mathrm{SMBH}$) and dimensionless accretion rates $\dot{m}(\equiv \dot{M}_\mathrm{acc}/\dot{M}_\mathrm{Edd})$,
{where $\dot{M}_\mathrm{acc}$ is the mass flow rate through the disk and $\dot{M}_\mathrm{Edd}$ is the Eddington mass accretion rate}.
We find that the radiative efficiency is mainly influenced by $\dot m$, while contributions of $M_\mathrm{SMBH}$ and $\dot{m}$ to the bolometric luminosity are comparably important.
We have developed new bolometric corrections linking the bolometric luminosity of an AGN to its luminosities in the hard X-ray, soft X-ray, and optical bands. Our results align with existing literature at high luminosities but suggest lower luminosities in the hard and soft X-ray bands at lower bolometric luminosities compared to those from commonly adopted empirical relations.
Combining with the semi-analytical model of galaxy formation \textsc{L-Galaxies} and Millennium dark matter simulation for the distribution of ($M_\mathrm{SMBH},\dot{m}$) at different redshifts, the model predicted AGN fraction shows good agreement with observational data to $z\lesssim2.5$, while the luminosity functions align well with observational data at $z\lesssim1$ but deviate from them for higher redshifts. This deviation may arise from improper treatments of SMBH growth at high redshifts in the galaxy formation model or bias from limited observational data. This AGN SED calculation can be readily applied to other cosmological simulations.

\end{abstract}



\section{Introduction} \label{sec:intro}

Active Galactic Nuclei (AGNs) are pivotal to understanding galaxy formation and evolution, highlighting the intricate relationship between supermassive black holes (SMBHs) and their host galaxies. Strong correlations between SMBH masses and bulge properties suggest a coevolution between black holes and classical bulges or elliptical galaxies. This connection was first proposed by \citet{dressler1989}, who identified a prominent relationship between SMBH mass and bulge luminosity. Subsequent studies have reinforced this finding, uncovering tight correlations between black hole mass and other bulge characteristics, including stellar velocity dispersion and bulge mass \citep{1998AJ....115.2285M,ferrarese2000, gebhardt2000, merritt2001, McLure2002, Haring&rix2004, kormendy2009, kormendy2013, 2024arXiv241100091P}. Moreover, recent studies have observed these correlations persisting at high redshifts \citep[e.g.,][]{2015ApJ...805...96S, 2019ApJ...887..245S}. Given the vast differences in mass and size between SMBHs and their host galaxies, these findings point to a complex interplay driving the growth of SMBHs alongside the evolution of their host galaxies.

The energy and momentum released during SMBH accretion could significantly impact the surrounding interstellar medium (ISM) by increasing gas temperatures and redistributing or expelling gas from the host galaxy \citep[e.g.,][]{springel2005,Vogelsberger2013,schaye2015}. 
This feedback mechanism could ultimately help to regulate star formation within galaxies. 
{Some observational studies show that central black holes could regulate cool gas accretion and star formation in massive galaxies \citep[e.g.,][]{2017ApJ...844..170T, 2022A&A...659A.160B, 2024Natur.632.1009W}, while others have revealed a more complex picture of AGN feedback \citep[e.g.,][]{2017NatAs...1E.165H, 2018ApJ...854..158S, 2019ApJ...873...90S, 2020ApJ...901...42Y, 2020ApJ...899..112S, 2021ApJ...906...38Z, 2022ApJ...935...72M}
suggesting that in some cases, AGN may promote star formation \citep[e.g.,][]{2015ApJ...799...82C, 2020ApJ...896..108Z, 2021ApJ...906...38Z} or show no clear relationship with star formation activities \citep[e.g.,][]{2013A&A...560A..72R, 2015ApJ...806..187A, 2015ApJ...805...96S, Shimizu_2016}.}

In most modern cosmological simulations, AGN feedback is now a standard feature, where it plays a central role in regulating star formation rates and shaping galaxy structures \citep{sijacki2007, Dubois2013, Vogelsberger2013, Vogelsberger2014, schaye2015, Crain2015, Choi2018}. In massive galaxies, AGN feedback is particularly critical for reproducing key observational features, such as suppressed star formation efficiency, and addressing the cooling flow problem \citep[e.g.,][]{2005MNRAS.361..776S, springel2005, Croton2006, Bower2006, Hlavacek-Larrondo2013, Gaspari2013, schaye2015, Yang2016, springel2018, 2023MNRAS.526.4978S, 2023MNRAS.524.2539P, 2023MNRAS.524.2556H, 2024Natur.632.1009W}. However, SMBH growth and AGN feedback processes occur on scales far smaller than those resolved by large-scale simulations of galaxy formation. For instance, the Schwarzschild radius of a $10^9\mathrm{M\mathrm{_{\odot}}}$ SMBH is approximately $\approx 10^{-4}$ pc. Consequently, these processes must be approximated using sub-grid models, which rely on assumptions and tunable parameters. These models are often calibrated phenomenologically to reproduce observed galaxy properties \citep[e.g.,][]{Vogelsberger2013,Vogelsberger2014, schaye2015, Henriques15}.



Constraints on the simulations rely on comparison between simulation predictions and observations. However, directly observing SMBH masses and accretion rates - the key outputs of simulations - remains challenging. 
{SMBH masses can be measured dynamically through the motion of gas, such as water masers \citep[e.g.,][]{1995Natur.373..127M, Gao2017}, ionized gas \citep[e.g.,][]{2001ApJ...555..685B}, or stars, as demonstrated in the Milky Way \citep[e.g.,][]{Ghez2003} and other galaxies \citep[e.g.,][]{2003ApJ...583...92G}. Yet, due to the small physical size of the sphere of influence of the SMBH, such data are only available for a few nearby galaxies \citep[see review, ][]{kormendy2013}. 
Reverberation mapping has extended SMBH mass estimates in active galaxies to higher redshift \citep[e.g.,][]{Shen2023}.}
Recently, the James Webb Space Telescope (JWST; \citealt{2023PASP..135f8001G}) has identified SMBHs at redshifts up to $z\approx10$ , with estimated masses in the range of $10^{6}-10^{8}\mathrm{M}_\odot$ \citep{Goulding2023, Bogdan2024, 2024ApJ...964...39G}.
However, these mass estimates, based on single-epoch spectroscopy of broad emission lines \citep[e.g.,][]{2005ApJ...630..122G} are subject to significant uncertainties, and the small sample sizes limit the statistical power of these observations.

Alternatively, simulated SMBH masses and accretion rates can be translated into observables for comparison with data. Advances in observational facilities have provided extensive datasets of AGN emissions across a wide range of wavelengths and cosmic epochs, from the nearby Universe \citep[e.g., the Sloan Digital Sky Survey, SDSS,][]{York2000} to quasars at redshifts as high as $z\approx6$ \citep{Fan2001, Fan2003, Fan2004, Jiang2009, Willott2010, Mortlock2011, Banados2018}. 
{More recently, JWST programs have identified numerous AGNs at redshifts $z \approx 3-12$ \citep{Juodzbalis2023, 2023ApJ...946L..14K, Goulding2023, Bogdan2024, 2024ApJ...964...39G}.}

However, the conversion of simulated SMBH masses and accretion rates into AGN luminosities is often oversimplified in cosmological simulations. For example, \citet{marulli2008,Henriques15, schaye2015, Weinberger2017} assume that the bolometric luminosity is proportional to the absolute mass accretion rate $L_\mathrm{bol} \propto \dot{M}_\mathrm{{acc}}$. \citet{hirschmann2014, Habouzit2021,Churazov2005} further modify the bolometric luminosity in the low accretion rate regime as $L_{\mathrm{bol}} \propto L_{\mathrm{Edd}} \times \dot{m}^2$. 
\citet{Vogelsberger2013} adopted an accretion-rate-dependent efficiency, with $L_\mathrm{bol}\propto\dot{m}^2 M_\mathrm{SMBH}$ at low accretion rates, and $L_\mathrm{bol}\propto\dot{m} M_\mathrm{SMBH}$ at high accretion rates. {\citet{Shirakata2019} adopted a more complex dependence on $\dot{m}$ as $L_\mathrm{bol}\propto [\frac{1}{1+3.5(1+\tanh{(\log{(\dot{m}/10)}))}}+\frac{10}{\dot{m}}]^{-1}$.}
\citet{Fanidakis2012} assumes an Advection-Dominated Accretion Flow (ADAF) for $\dot{m}<0.01$ \citep{mahadevan1997}, a standard thin disk for $\dot{m}>0.01$ \citep{SSD1973}, and a radiative efficient standard disk at super-Eddington regime \citep{SSD1973}.
Similarly, \citet{griffin19} split the accretion flow into three regimes based on the accretion rate, adopting the slim disk model, standard disk model, and ADAF for decreasing accretion rate respectively.
However, the efficiency could be more complex than these assumptions suggest. For example, \citet{2012MNRAS.427.1580X} studied the radiative efficiency of the hot accretion flow at both low and high accretion rates, revealing a complex increasing trend in radiation efficiency with accretion rate. Such radiation efficiency was subsequently incorporated into two-dimensional hydrodynamical numerical simulations to investigate the connection between AGNs and their host galaxies \citep{2018ApJ...857..121Y}.


Moreover, earlier studies often rely on bolometric corrections derived from typical SEDs to convert bolometric luminosity into photometric measurements in specific bands for comparison with observations  \citep[e.g.,][]{marconi2004, 2020MNRAS.495.3252S}. However, these bolometric corrections frequently overlook the variation in AGN SEDs and their associated physical properties. For instance, the peak frequency of the spectrum can shift to higher values at increased accretion rates for a given bolometric luminosity \citep[e.g.,][]{2017MNRAS.465..358C}. Therefore, a more sophisticated model is required to accurately represent the diverse shapes of AGN SEDs. 

In this study, we develop a model to calculate the emergent spectrum of AGN  based on the mass of the SMBH and its Eddington-normalized accretion rate. 
 We account for two distinct accretion regimes based on the Eddington-normalized accretion rate: the modified magnetic reconnection-heated disk corona model for high-accretion-rate objects and the ADAF+thin disk model for low-accretion-rate objects. Using the complete SED, we can derive the luminosity in any chosen filter by integrating over the corresponding frequency range. By incorporating the SMBH mass and accretion rate from a semi-analytical galaxy catalog, this approach enables more accurate comparisons with observational data.

The paper is organized as follows. In Section.\,\ref{sec:simulation}, we provide a brief description of the cosmological simulation and semi-analytical galaxy formation models used in our study, as well as the modifications; in Section.\,\ref{sec:model}, we introduce SMBH SED models as a function of SMBH mass and accretion rate and discuss the radiative efficiencies; We employ the AGN SED model in post-processing, leveraging black hole masses and accretion rates predicted by the semi-analytic galaxy models to examine AGN observable properties in Section.\,\ref{sec:lum}. We present our main results, including the predicted optical-X-ray spectra index, new bolometric corrections between AGN bolometric luminosity and luminosities in hard X-ray, soft X-ray, and optical bands, AGN fraction as a function of redshift, and AGN luminosity functions. 
Section.\,\ref{sec:conclusion} summarizes our result.
In Appendices, we illustrate the intrinsic variation of AGN SEDs, discuss model parameter influences, detail modifications to the slim disk component, and present comparisons between model predictions with observational data for individual sources. 

\section{Galaxy formation and SMBH growth} \label{sec:simulation}
\subsection{N-body simulations and semi-analytical galaxy formation models}
The semi-analytical galaxy catalog is generated by implementing galaxy formation models \textsc{L-Galaxies} \citep[e.g.,][]{Springel2001,Croton2006,Guo2011} onto the dark halo merger tree from the rescaled Millennium $N$-body cosmological simulation \citep{springel2005}. The Millennium simulation adopts cosmological parameters consistent with first-year WMAP cosmological parameters, which are then rescaled to match the Planck result \citep{Angulo2010} with the following parameters: $\sigma_8=0.829, h=0.673, \Omega_\Lambda=0.685, \Omega_m=0.315, \Omega_b=0.0487$. It traces $2160^3$ particles from redshift 57.6 to the present day in a cubic box with a side length of 480.279 Mpc$\,h^{-1}$. The particle mass resolution is 9.61 $\times 10^8\,\mathrm{M_\odot}\,h^{-1}$. The particle data is stored at 64 snapshot outputs. Dark halos are identified using the standard friend-of-friend (FOF) method, where particles with a separation less than 0.2 times the averaged interparticle separation \citep{Davis1985} are linked together. The SUBFIND algorithm \citep{Springel2001} is then applied to identify self-bound substructures/subhalos. Merger trees are then built up by linking each subhalo to its unique descendant. For more details about the halo and subhalo finder and the construction of the merger trees, please refer to \citet{springel2005}.


{\textsc{L-Galaxies} models the baryonic processes along the dark halo merger tree in a semi-analytical way. A fraction of baryons fall into the collapsed dark matter halo and become shock-heated. These baryons either cool rapidly onto the galactic disk on a free-fall timescale or form a hot atmosphere that gradually cools and flows toward the potential well, a process known as a cooling flow. The inflowing cold gas is assumed to have the same specific angular momentum as the dark matter, determining the radius of the gas disk. Star formation occurs in the cold gas disk, with stellar evolution processes returning energy, mass, and heavy elements to the surrounding medium. This feedback reheats the cold gas, transferring it back into the hot atmosphere and, in some cases, ejecting it into an external reservoir outside the halo. The ejected gas can later re-accrete into the halo. 

Environmental effects on satellite galaxies are also incorporated. Upon crossing the virial radius of the primary halo, tidal forces act to strip hot gas, cold gas, and stars from the satellite, while ram pressure removes hot gas in massive halos. These processes could be effective in quenching star formation in satellite systems. Galaxy mergers, which trigger bulge/elliptical galaxy formation and starburst activity, are assumed to follow the mergers of their host dark matter halos, with some time delay. Bulges can also form through disk instabilities. 

SMBHs in the potential centre grow through black hole mergers and the accretion of cold and hot gas. During gas accretion, a large amount of energy is released, preventing the hot gas in the intracluster medium (ICM) from cooling in clusters and groups. This, in turn, reduces star formation rates, particularly in massive systems.

In this study, we start with the recent version of the \textsc{L-Galaxies} model developed by \citet{Henriques15} (hereafter \citetalias{Henriques15}) and briefly outline the prescriptions for modeling SMBH growth.}

\begin{table*}
\begin{center}
\caption{{\bf Best-fit parameters for L-Galaxies and parameters for the AGN SED model} The best-fit values in L-Galaxies are compared with those in  \citetalias{Henriques15}.
The last row presents the model parameters of the AGN SED model used in this work, including the ADAF+thin disk model and the magnetic reconnection-heated disk-corona model. Here $\alpha$ represents the viscosity parameter, $\beta[\equiv p_\mathrm{g}/(p_\mathrm{g}+p_\mathrm{m})]$ is the magnetic parameter, $\delta$ denotes the fraction of the viscously dissipated energy that directly heats the electrons in ADAF, and $\beta'[\equiv p_\mathrm{g}/p_\mathrm{m}]$ is the magnetic parameter specific to the disk-corona model.}
\label{tab:para}
\begin{tabular}{lcccc}
\hline
\hline
\multicolumn{2}{c}{}&this work &Henriques15 &Units\\
\hline
\hline
\multicolumn{2}{c}{$\alpha_{\rm{SF}}$ (SF eff)} & 0.028 &0.025\\
\multicolumn{2}{c}{$\Sigma_{\rm{SF}}$ (Gas density threshold)} & 0.11 &0.24 &$\rm{10^{10}\mathrm{M}_\odot\,pc^{-2}}$\\
\multicolumn{2}{c}{$\alpha_{\rm{SF, burst}}$ (SF burst eff)} & 0.49 &0.60 \\
\multicolumn{2}{c}{$\beta_{\rm{SF, burst}}$ (SF burst slope)} & 1.9 &1.9 \\
\hline
\hline
\multicolumn{2}{c}{$k_{\rm{AGN}}$ (Radio feedback eff)} &$1.1\times 10^{-3}$ &$5.3\times10^{-3}$ & $\rm{\mathrm{M}_\odot\,yr^{-1}}$\\
\multicolumn{2}{c}{$f_{\rm{BH}}$ (BH growth eff)} & 0.091 &0.041 \\ 
\multicolumn{2}{c}{$V_{\rm{BH}}$ (Quasar growth scale)} & 547 &750 &$\rm{km\,s^{-1}}$\\ 
\hline
\multicolumn{2}{c}{$\epsilon$ (Mass-loading eff)} & 0.8 &2.6  \\
\multicolumn{2}{c}{$V_{\rm{reheat}}$ (Mass-loading scale)} & 515 &480  &$\rm{km\,s^{-1}}$\\
\multicolumn{2}{c}{$\beta_{1}$ (Mass-loading slope)} & 1.11 &0.72  \\
\multicolumn{2}{c}{$\eta_\mathrm{SN}$ (SN ejection eff)} & 1.18 &0.62 \\
\multicolumn{2}{c}{$V_{\rm{eject}}$ (SN ejection scale)} & 131 &100  &$\rm{km\,s^{-1}}$\\
\multicolumn{2}{c}{$\beta_{2}$ (SN ejection slope)} & 0.97 &0.80 \\
\multicolumn{2}{c}{$\gamma$ (Ejecta reincorporation)}
&$4.3\times10^{10}$ &$3.0\times10^{10}$   & yr\\
\hline
\multicolumn{2}{c}{$M_{\rm{r.p.}}$ (Ram-pressure threshold)} &$1.2\times10^{4}$ &$1.2\times10^{4}$ &$\rm{10^{10}\mathrm{M}_\odot}$\\
\multicolumn{2}{c}{$R_{\rm{merger}}$ (Major-merger threshold)} & 0.1 &0.1 \\
\multicolumn{2}{c}{$\alpha_{\rm{friction}}$ (Dynamical friction)} & 0.91 &2.5 \\
\hline
\multicolumn{2}{c}{$y$ (Metal yield)} & 0.039 &0.046 \\
\hline
\hline
\multirow{2}{*}{Accretion model parameters} & $\alpha$ & $\beta$ & $\delta$ & $\beta'$\\\cline{2-5}
& 0.05 & 0.95 & 0.2 & 8 \\
\hline
\hline
\end{tabular}
\end{center}
\end{table*}

\subsection{SMBH growth}\label{sec:SMBH_growth}

In \citetalias{Henriques15}, each galaxy hosts an SMBH seed with zero initial mass at the potential center when its host halos are first identified. SMBHs grow through two primary modes: quasar-mode and radio-mode, which are described as follows. This quasar-mode is initiated by galaxy mergers, which channel a significant amount of cold gas toward the central region, fueling the SMBH. The amount of gas accreted during the time interval between two consecutive snapshots when a merger occurs, $\Delta M_{\mathrm{SMBH}}$, depends on the specific characteristics of the merging galaxies. We use the average value of  $\Delta M_{\mathrm{SMBH}}/ \delta t_\mathrm{snapinterval}$ as the absolute SMBH mass accretion rate.
\begin{equation}
\Delta M_{\mathrm{SMBH}, \mathrm{Q}}=\frac{f_{\mathrm{SMBH}}\left(M_{\mathrm{sat}} / M_{\mathrm{cen}}\right) M_{\mathrm{cold}}}{1+\left(V_{\mathrm{SMBH}} / V_{200 \mathrm{c}}\right)^2},
\end{equation}
where $M_\mathrm{cen}$ and $M_\mathrm{sat}$ represent the total baryon masses of the central and satellite galaxies.  $M_\mathrm{cold}$ represents their total cold gas mass, $V_\mathrm{200c}$ is the viral velocity of the central halo, and $f_\mathrm{SMBH}$ and $V_\mathrm{SMBH}$ are two adjustable parameters that govern the fraction of available cold gas accreted. After the galaxies merge, the final mass of the black hole ($M_\mathrm{SMBH,f}$) is obtained by summing up the masses of the individual black holes prior to the merger and the accreted gas during mergers,
$M_\mathrm{SMBH,f}=M_\mathrm{SMBH,1}+M_\mathrm{SMBH,2}+\Delta M_{\mathrm{SMBH,Q}}$. The majority of black hole mass growth occurs during the quasar-mode \citep[e.g.,][]{Zhang2021}.

The radio-mode of accretion is characterized by a relatively low accretion rate, where the black hole continuously captures gas from the hot atmosphere surrounding its host galaxy. In this scenario, a phenomenological accretion model is often assumed \citep{Croton2006}, which describes the accretion process using the host galaxy hot gas mass and SMBH mass:
\begin{equation}
\dot{M}_{\mathrm{SMBH}}=k_{\mathrm{AGN}}\left(\frac{M_{\mathrm{hot}}}{10^{11} \mathrm{M}_{\odot}}\right)\left(\frac{{M}_{\mathrm{SMBH}}}{10^8 \mathrm{M}_{\odot}}\right) .
\end{equation}
where $k_\mathrm{AGN}$ is the AGN radio-mode efficiency, $M_\mathrm{hot}$ is the hot gas mass, and $M_\mathrm{SMBH}$ is the SMBH mass.


The Eddington-luminosity $L_\mathrm{Edd}$ refers to the maximum luminosity of a steady, spherically symmetric accretion flow, at which the outward radiative pressure balances out the inward gravitational force. It is related to the Eddington mass by 
$L_\mathrm{Edd}= 0.1\dot{M}_\mathrm{Edd}$. 
In reality, the coefficient may depend on SMBH properties and accretion rate. Here we adopt $\eta=0.1$, a commonly used value in the literature. The SMBH accretion rate is commonly expressed as the Eddington-normalized form $\dot{m}=\dot{M}_\mathrm{acc}/\dot{M}_\mathrm{Edd}$, where $\dot{M}_\mathrm{acc}$ is the mass accretion rate in g\,s$^{-1}$.
We use this definition in the following sections unless stated otherwise.


\begin{figure}
\includegraphics[width=\columnwidth]{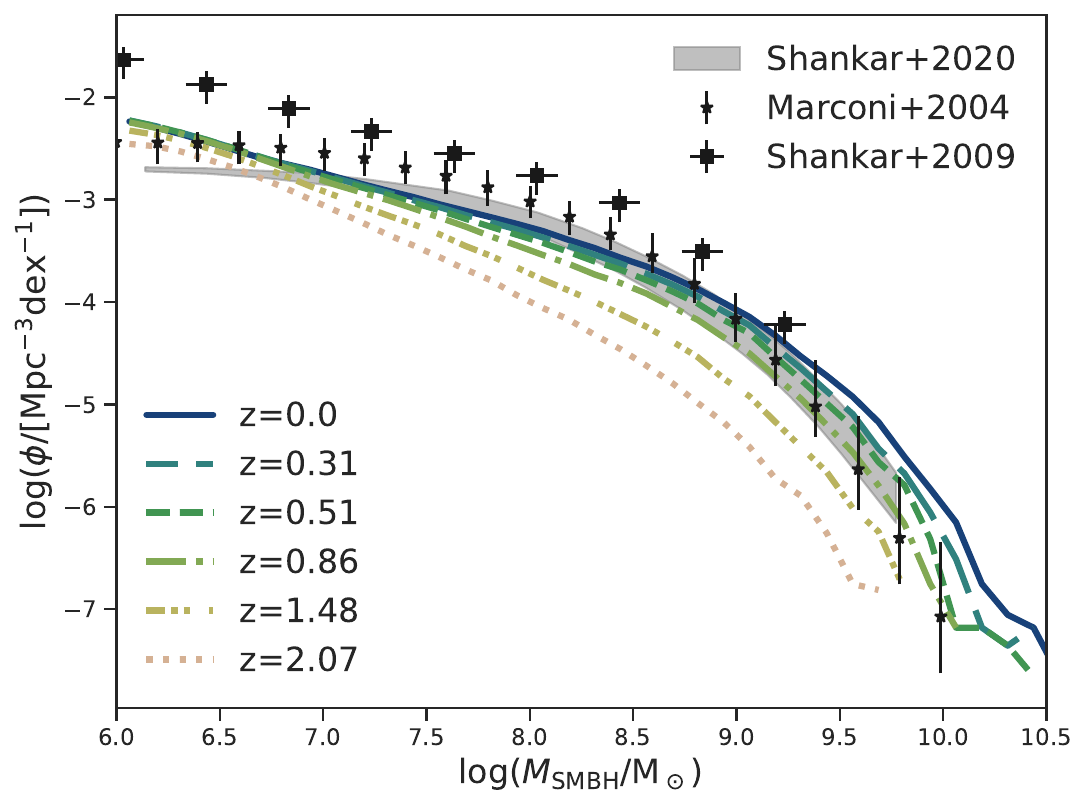}
\caption{\textbf{Black hole mass functions.} Solid curves in different colors represent the model predictions at various redshifts, with the corresponding redshift values indicated in the lower-left corner. Observational data at $z = 0$ are taken from \citet{marconi2004} (stars), \citet{2009ApJ...690...20S}(squares), and \citet{2020NatAs...4..282S} (shaded area).
\label{fig:BHMF}}
\end{figure}

\begin{figure}[h!]
\centering
\includegraphics[width=\columnwidth]{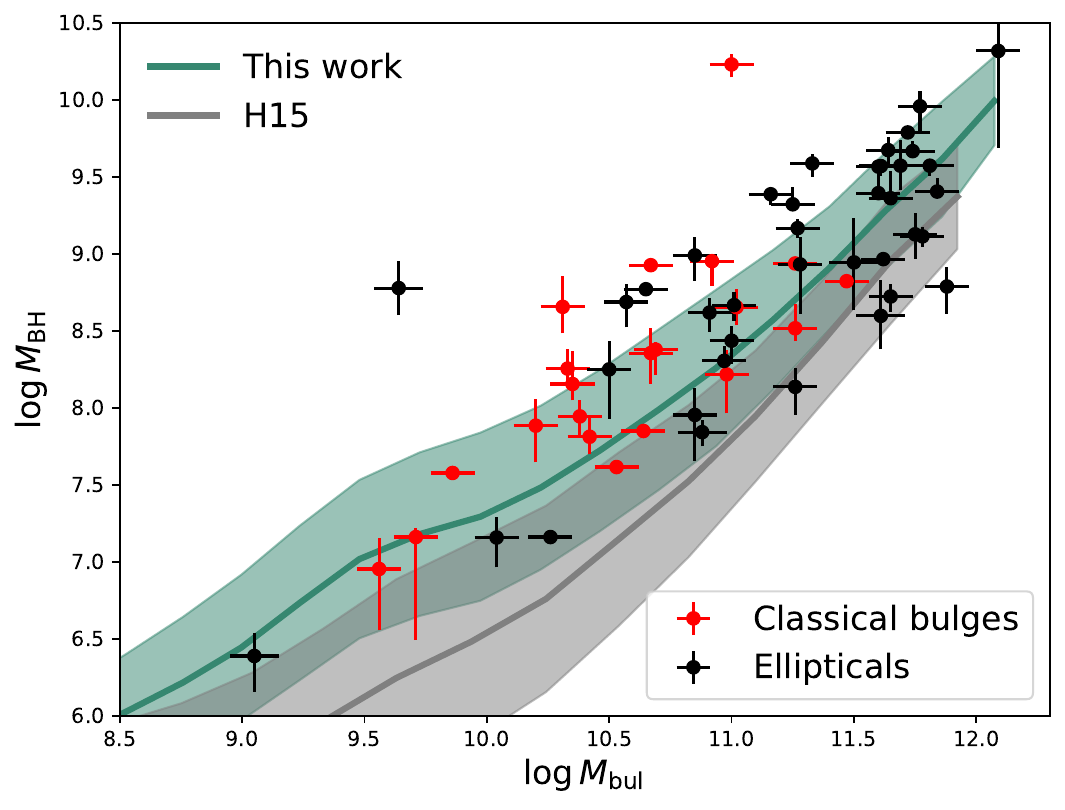}
\caption{\textbf{Comparison of the $M_\mathrm{BH}-M_\mathrm{bul}$ relations predicted by the semi-analytical catalog adopted in this work and the H15 catalog}. 
The solid curves denote the median BH mass in each bulge mass bin in the two simulated catalogs, with the shaded areas denoting the $\pm1\sigma$ value. The red and black scatter points show the observed $M_\mathrm{BH}-M_\mathrm{bul}$ pairs presented in Table.\,3 and Table.\,2 of \citet{kormendy2013}, each corresponds to galaxies with classical bulges and ellipticals, excluding those sources that lack one of the two estimated values and those marked as merger in progress.
\label{fig:MbhMbul}}
\end{figure}

\begin{figure*}
\includegraphics[width=2\columnwidth]{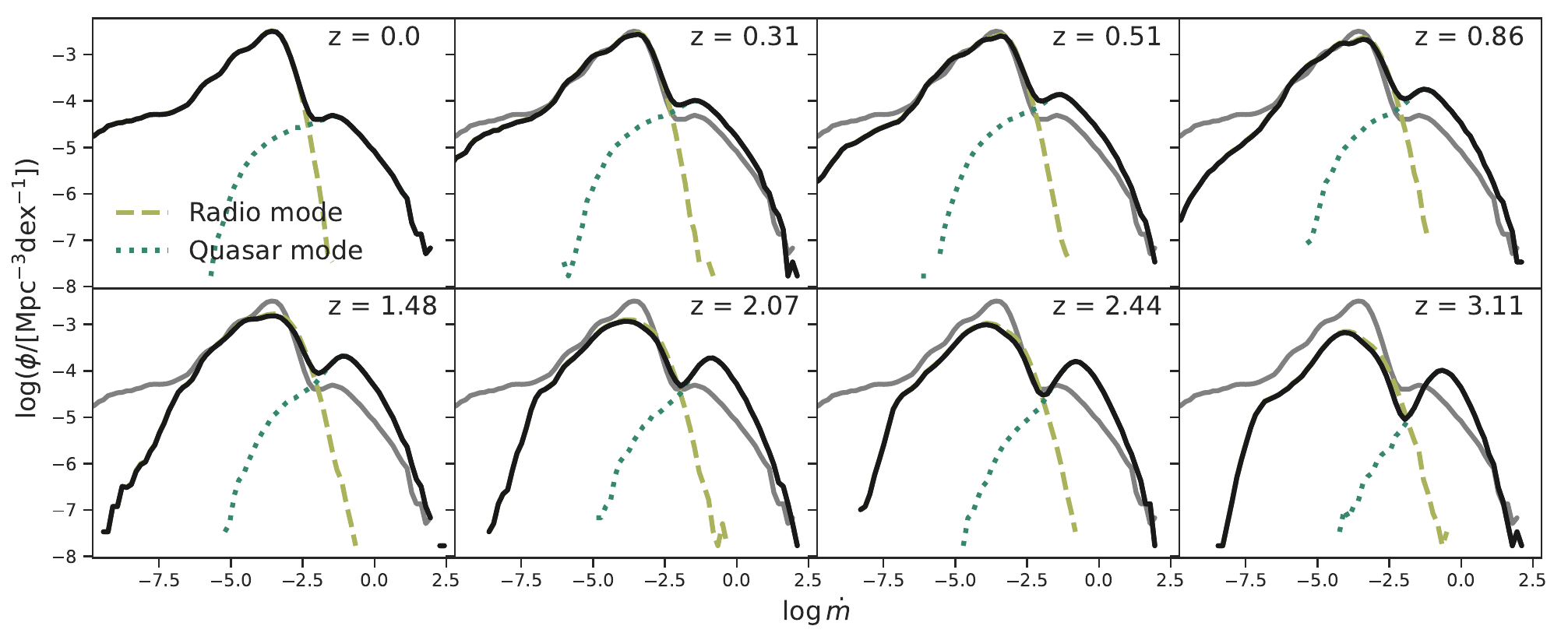}
\caption{
\textbf{ Distribution of Eddington-normalized SMBH accretion rates $\log \dot{m}$ at different redshifts.} The black curves show the overall distributions for SMBHs with mass $\log(M_\mathrm{SMBH}/\mathrm{M_\odot})>6$, while the colored curves indicate the contributions from radio-mode and quasar-mode accretion, with the color coding shown in the first panel. The grey curve in each panel duplicates the $\log \dot{m}$ distribution at $z=0$ for comparison. \label{fig:mdot}}
\end{figure*}

\begin{figure*}
\includegraphics[width=2\columnwidth]{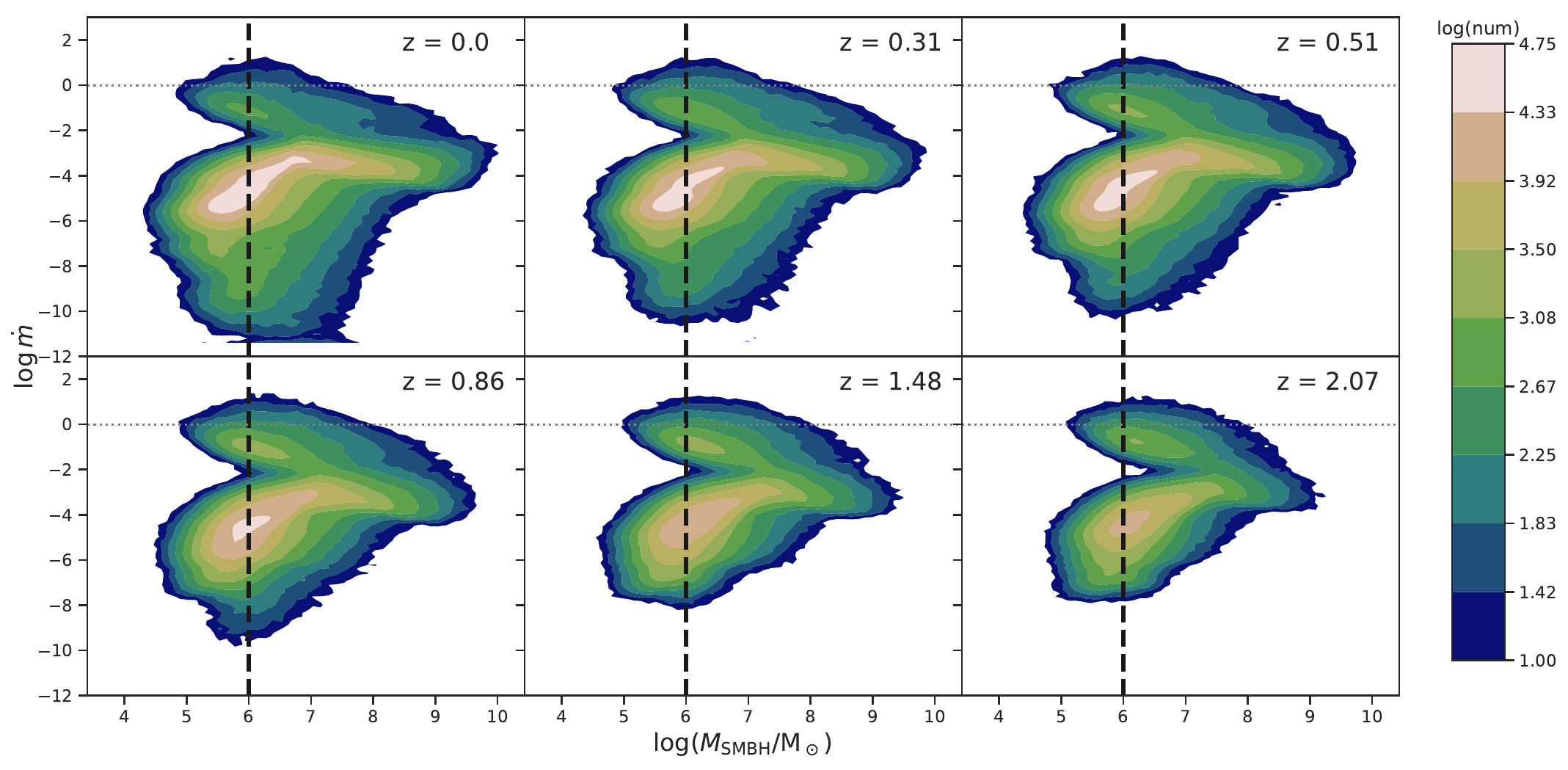}
\caption{
\label{fig:mass-BHAR} \textbf{Joint distribution of SMBH mass, $\log (M_\mathrm{SMBH}/\mathrm{M_\odot})$, and SMBH accretion rate, $\log{\dot{m}}$.} Results at redshifts $z=0, 0.31, 0.51, 0.86, 1.48, 2.07$ are displayed in separate panels as in the top right corner. In each redshift, the distributions are color-coded based on the logarithm of the SMBH number in each bin in the $\dot{m}-M_\mathrm{SMBH}$ plane. 
The black dashed lines denote $\log(M_\mathrm{SMBH}/\mathrm{M_\odot})=6$, below which the SMBHs could suffer from resolution effects. For reference, the gray dotted lines denote the Eddington accretion rate.
}
\end{figure*}

\subsection{Semi-analytical model parameters}

It is found that \citetalias{Henriques15} fails to reproduce the observed SMBH mass vs. spheroid mass relation, with the predicted SMBH being an order of magnitude lower than the observed values at $z=0$ \citep[e.g.,][]{2024MNRAS.529.4958P}. We employ Markov Chain Monte Carlo (MCMC) to calibrate the semi-analytic model parameters. In addition to the previously adopted constraints, i.e., the stellar mass function at $z = 0-3$ \citep{2008MNRAS.388..945B, 2009MNRAS.398.2177L, 2009ApJ...701.1765M, 2010ApJ...709..644I, 2010ApJ...725.1277M, 2011MNRAS.417..900D, 2012MNRAS.421..621B,  2013A&A...556A..55I, 2013ApJ...777...18M, 2014ApJ...783...85T} and red galaxy fraction at $z = 0-3$ \citep{2003ApJS..149..289B, 2004ApJ...600..681B, 2013A&A...556A..55I, 2013ApJ...777...18M, 2014ApJ...783...85T}, we also incorporate the local SMBH mass function at $z=0$ \citep{2009ApJ...690...20S}. The relevant parameters and their comparison with \citetalias{Henriques15} are presented in Table.\,\ref{tab:para}. 
Here $\alpha_{\rm{SF}}$ is the star formation efficiency and $\Sigma_{\rm{SF}}$ is the cold gas density threshold above which star formation could occur. The parameters $\alpha_{\rm{SF, burst}}$ and $\beta_{\rm{SF, burst}}$ control star formation during mergers, following the relationship of $\delta M_\mathrm{*,starburst}=\alpha_\mathrm{SF,burst}(M_1/M_2)^{\beta_\mathrm{SF,burst}} M_\mathrm{cold}$ \citep{2001MNRAS.320..504S}, where $\delta M_\mathrm{*,starburst}$  is the stellar mass formed during collisional starburst, $M_1$ and $M_2$ are the baryonic masses of the two merging galaxies, and $M_\mathrm{cold}$ is their total cold gas mass. $k_{\rm{AGN}}$ is the efficiency of radio accretion while $f_{\rm{BH}}$ and $V_{\rm{BH}}$ regulate the quasar accretion rate. $\epsilon$, $V_{\rm{reheat}}$ and $\beta_{1}$ determine the reheated cold gas mass during SN feedback, while $\eta_\mathrm{SN}$, $V_{\rm{eject}}$ and $\beta_{2}$ determine the total energy released by SN feedback. $\gamma$ adjusts the time scale of the reincorporation of ejected gas. $M_{\rm{r.p.}}$ is the main halo mass threshold above which ram pressure is started for satellites. $R_{\rm{merger}}$ is the threshold for major and minor mergers, which we fixed at 0.1, and $\alpha_{\rm{friction}}$ multiplies the dynamical friction delay for mergers. $y$ is the total mass of metals produced by each solar mass star. More details about the model parameters can be found in \citetalias{Henriques15}.

\subsection{SMBH mass functions, scaling relation and accretion rates}

Before analyzing AGN luminosities, we first examine the SMBH mass function and its evolution, as well as their accretion rates. In practice, at each \textbf{redshift snapshot} of interest, all SMBH masses in the simulation are collected, binned in logarithmic intervals of 0.125 dex, and normalized by the comoving volume to yield the SMBH mass function.
Fig.\,\ref{fig:BHMF} shows the SMBH mass function and its variation with redshift. At $z=0$, the model prediction aligns broadly well with the observed data. Toward higher redshifts, the abundance decreases, and the slope steepens at lower masses.
It should be noted that the abundance of SMBHs with masses below $10^6\,\mathrm{M_\odot}$ could suffer from halo mass resolution effects. Therefore, our main analysis focuses on SMBHs with masses $>10^6\,\mathrm{M_\odot}$.

Fig.\,\ref{fig:MbhMbul} compares the $M_\mathrm{BH}-M_\mathrm{bul}$ relations predicted by the SAM used in this work and by the H15 catalog, where the solid curves represent the median black hole mass in each bulge mass bin, with shaded regions indicating the $\pm1\sigma$ scatter. The red and black data points show the observed $M_\mathrm{BH}-M_\mathrm{bul}$ pairs compiled in Table.\,2 and Table.\,3 of \citet{kormendy2013}, corresponding to galaxies with classical bulges and ellipticals. We exclude sources lacking one of the two estimates and those identified as mergers in progress.
Compared with the observed relation, for a given bulge mass, the predicted BH mass in the H15 catalog falls below the observational relation (also see Fig.\,9 of \citet{2024MNRAS.529.4958P}). In contrast, the SAM catalog employed in this study yields a relation that more closely matches the observed trend, both in normalization and shape (Note that the $M_\mathrm{BH}-M_\mathrm{bul}$ relation had not been adopted as calibration in the SAM catalog).

Fig.\,\ref{fig:mdot} illustrates the number density distribution of the Eddington-normalized accretion rate at different redshifts, showing in black solid curves. The accretion rates are further divided into quasar-mode and radio-mode accretion, denoted by dotted and dashed curves respectively, where quasar-mode predominantly drives moderate to super-Eddington rates, and radio-mode is associated with lower rates.
In general, the Eddington-normalized accretion rate distribution exhibits a double-peaked structure, with the higher-rate peak dominated by quasar-mode accretion. The $z=0$ accretion rate number density is duplicated in each panel for reference, denoted by grey solid curves. Compared to the number density distribution at each redshift, the high-accretion-rate peak becomes increasingly prominent at higher redshifts, while AGNs with very low accretion rates exist predominantly at low redshifts - accretion rates below $10^{-8}$ Eddington only present at $z<1.5$. This indicates that the contribution from quasar-mode accretion rises with redshift, reflecting the higher rate of mergers at earlier times.

Fig.\,\ref{fig:mass-BHAR} shows the joint distribution of SMBH mass and accretion rate across redshifts, revealing a bimodal pattern in the density map for $M_\mathrm{SMBH}<10^8\,\mathrm{M_\odot}$ across all the redshifts of interest. The vertical dashed curves represent the resolution limit at $M_{\rm SMBH} \approx 10^6\,\mathrm{M_{\odot}}$, and the gray dotted lines denote the Eddington accretion rate. One peak occurs around 10$^{-4}$ Eddington accretion rate, and the other near the Eddington limit with a separation point around 10$^{-2}$. We note that the exact value of the peak masses could be affected by the incomplete SMBH samples at masses lower than $10^6 \mathrm{M_{\odot}}$. The proportion of AGNs with high accretion rates around the high peak increases with redshift. At lower accretion rates, the scatter decreases with increasing redshifts, indicating that extremely low accretion rates ($<10^{-8}$) only occur at low redshifts.

For high-mass SMBHs ($M_\mathrm{SMBH}>10^8\,\mathrm{M_\odot}$), most AGNs accrete at a low accretion rate below Eddington. The scatter in accretion rates is smaller for high masses than for low masses. Notably, at $z=0$, about $80$\% 
of massive SMBHs ($>10^{8.5}\mathrm{M_{\odot}}$) exhibit accretion rates between [$10^{-4.2}, 10^{-2.9}$] with a minimum near $10^{-5}$ Eddington accretion rate. Similar narrow accretion distributions are observed at higher redshifts, despite the increased frequency of mergers. 

In summary, the SMBH mass function agrees with observed data at $z=0$, where constraints are most reliable. SMBHs show a bimodal distribution in accretion rates, with greater variability for lower-mass SMBHs. In contrast, high-mass SMBHs exhibit less variation, typically remaining below the Eddington accretion rate.

\section{The accretion modes and the emergent spectra in AGNs}\label{sec:model}

\begin{figure}
\includegraphics[width=\columnwidth]{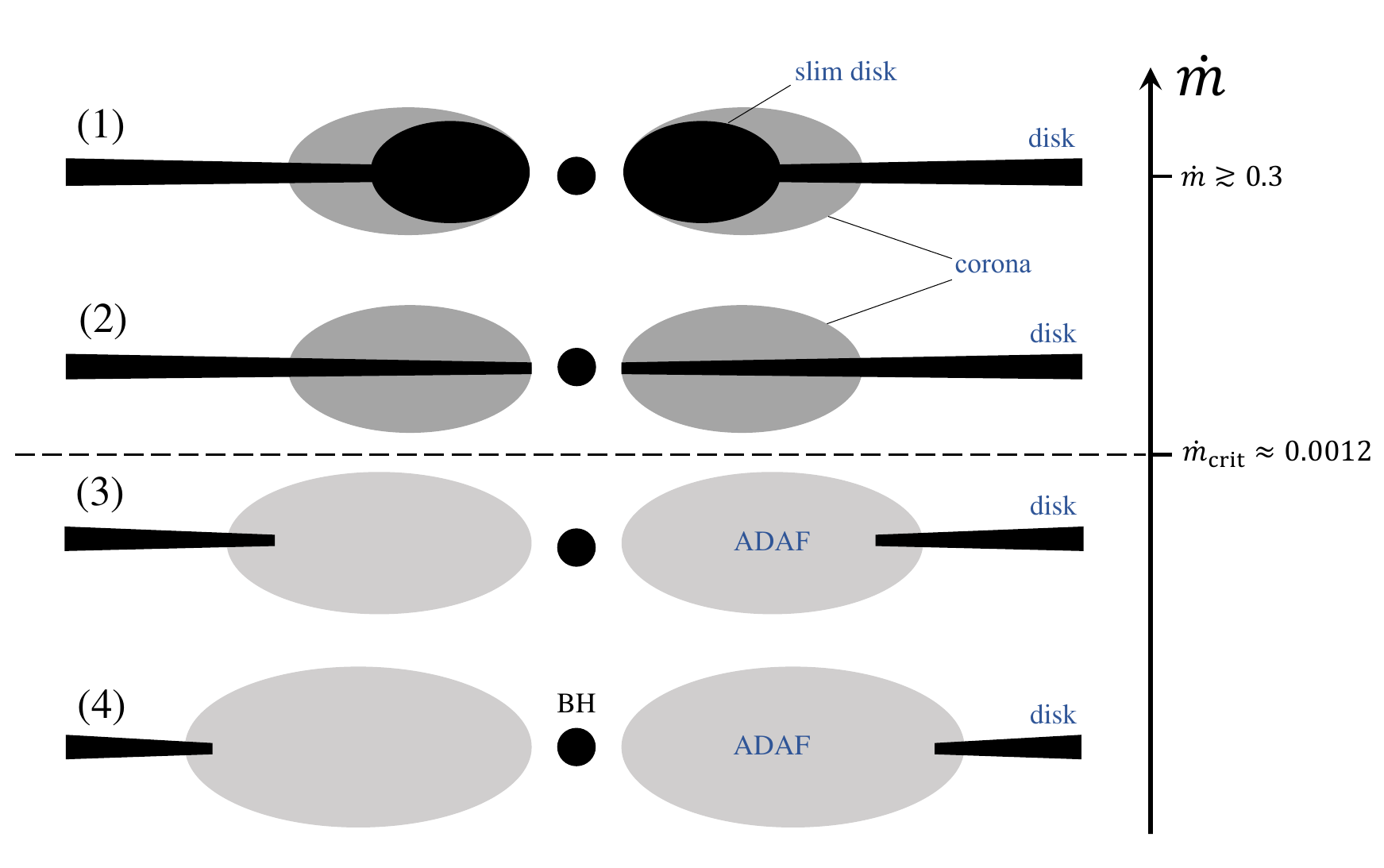}
\caption{\textbf{Schematic diagram of the accretion flow geometry at different accretion rates $\dot m$.} 
The critical mass accretion rate $\dot m_{\rm crit}$ defines the transition between the ADAF+disk and disk-corona accretion regimes. For $\dot{m}<\dot m_{\rm crit}$ (regions 3, 4), the accretion flow consists of an inner ADAF and an outer truncated accretion disk, with the disk truncation radius decreasing as $\dot m$ increases. For $\dot{m}>\dot m_{\rm crit}$ (regions 1, 2), the flow transitions to a disk-corona configuration, where the accretion disk extends down to the SMBH's ISCO and is surrounded by a hot corona. At high accretion rates (region 1) with $\dot m \gtrsim 0.3$, the corona in the innermost regions nearly collapses, causing vertical inflation of the disk to form a slim disk. The light grey area corresponds to the ADAF regime, the thin black region represents the thin disk, dark grey indicates the hot corona phase, and the black elliptical region represents the slim disk. 
\label{fig:geometry}}
\end{figure}
\begin{figure}
\includegraphics[width=\columnwidth]{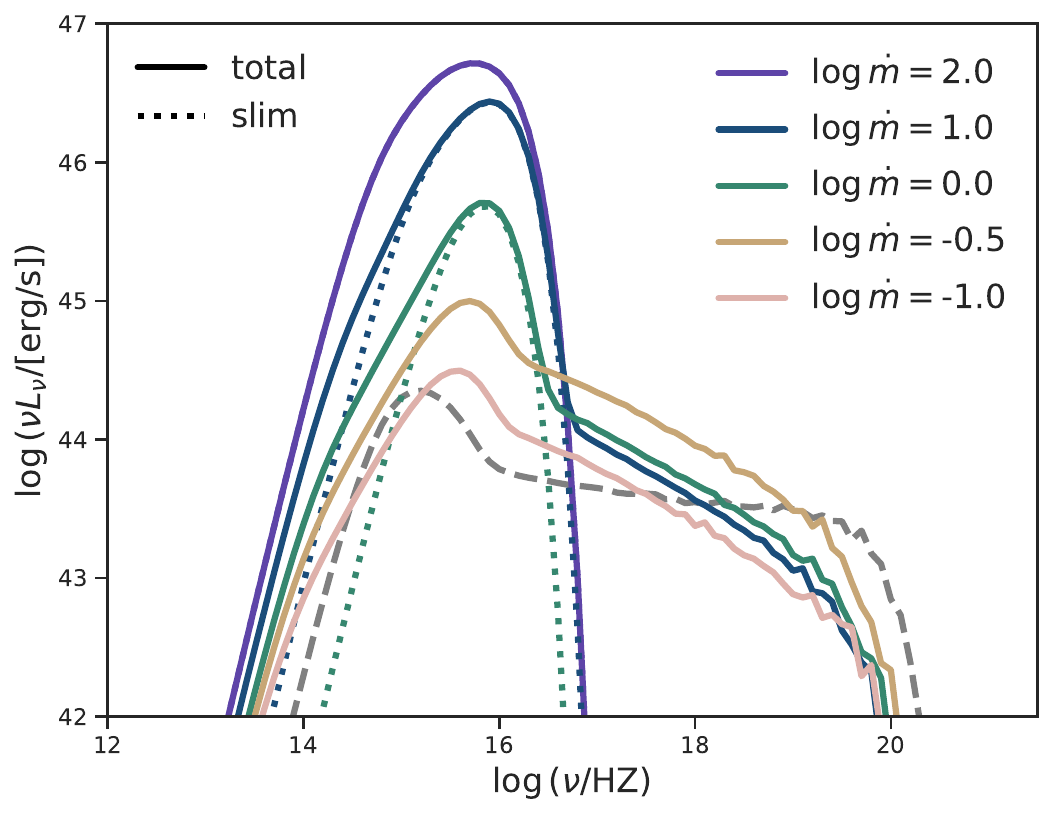}
\caption{\textbf{SEDs of the disk-corona model of a $10^{8} \,\mathrm{M}_\odot$ SMBH.} The SEDs are calculated using the magnetic reconnection-heated disk corona model for accretion rates ranging from$\log{\dot{m}} = -1.0$ to $2.0$. Solid curves represent the total SED, while dotted curves show the increasing contribution of the inner slim-like region with higher accretion rates. For comparison, the SED corresponding to $\log{\dot{m}}=-1.0$ from \citet{liuBF2003} is included as a grey dashed curve. 
\label{fig:SED_diskcor}}
\end{figure}

AGNs are believed to be powered by accretion onto the central SMBHs. The study of the specific accretion mode for different types of AGNs has been ongoing since the discovery of AGNs in the 1960s \citep[][for review]{2008ARA&A..46..475H,Netzer2015}. 
Based on their Eddington-ratio ($L_\mathrm{bol}/L_\mathrm{Edd}$), AGNs are divided into high-luminosity AGNs (HLAGNs) and low-luminosity AGNs (LLAGNs). 
Observationally, great efforts have been made to construct the simultaneous SED of AGNs, with which the bolometric corrections could be derived (e.g., X-rays and H$\alpha$ luminosity, \citealt{2009ApJ...699..626H}, for LLAGN; X-ray luminosity, \citealt{2007MNRAS.381.1235V, 2009MNRAS.399.1553V}, for HLAGN). 

In general, the accretion flow of LLAGNs is suggested to have a geometry with two components, i.e., an inner advection-dominated accretion flow (ADAF) and an outer truncated thin disk \citep[e.g.,][]{2019ApJ...870...73Y}. In this configuration, the ADAF dominates the X-ray emission, and depending on the value of the truncation radius, the truncated thin disk dominates the optical to infrared \citep[][for review]{Quataert1999, Yuan2004, Nemmen2014, Lasota1996, Nemmen2006, YuZ2011, WuYanYi2013, Storchi-Bergmann2003, Eracleous2009, 2008ARA&A..46..475H}.
For high-luminosity AGNs, the accretion flow is suggested to have a geometry with the thin accretion disk sandwiched by a hot corona extending down to the innermost stable circular orbits (ISCO) of the SMBH, in which the accretion disk dominates the optical and UV emission, and the corona dominates the X-ray emission \citep[][]{SSD1973, Shields1978, Malkan1982, Mushotzky1993, Shang2005, Haardt1991, Haardt1993, Nakamura1993, Svensson1994, Dove1997, Kawaguchi2001}.
The geometry evolution of the accretion flow in different types of AGNs is very similar to the spectral state evolution found in BH low-mass X-ray binaries, in which the spectral state evolution is dominantly determined by $\dot m$ \citep[][]{Esin1997}. Great efforts have been made to establish the intrinsic connection of the accretion flow geometry between BH X-ray binaries and AGNs \citep[][]{Merloni2003, Wuqingwen2008, Gallo2018, Qian2018, Gultekin2019, Arcodia2020, Bariuan2022}.

In this study, we construct a scenario for the accretion flow evolution in different types of AGNs across various $\dot m$, as shown in Fig.\,\ref{fig:geometry}.
In our model, there exists a critical Eddington-normalized accretion rate, i.e., $\dot m_{\rm crit}$, between region (2) and region (3). Below $\dot m_{\rm crit}$, the geometry of the accretion flow is an inner ADAF plus an outer truncated thin accretion disk; Above $\dot m_{\rm crit}$, the geometry of the accretion flow is an accretion disk (or a slim disk for higher $\dot m$) extending to the ISCO of the BH, sandwiched by a hot corona. Observationally, $\dot m_{\rm crit}$ is constrained to be $\approx1\%$ or less \citep{2008ARA&A..46..475H}. Several models have been proposed to explain the observed value of $\dot m_{\rm crit}$, among which the disk evaporation model is believed to be one of the most promising. 
Additionally, the disk evaporation model can also effectively explain the observed truncation radius of the accretion disk for $\dot m \lesssim \dot m_{\rm crit}$ in low-luminosity AGNs (See Appendix.\,\ref{sec:SED4panel} for the example SED of each state).
In the following, we summarize the properties of the accretion flow in the framework of the disk evaporation model for $\dot m \lesssim \dot m_{\rm crit}$, and the disk-corona model with the corona built via magnetic reconnection for $\dot m \gtrsim \dot m_{\rm crit}$ respectively. Note that SMBH spin is not considered in the present paper. 

\subsection{The disk evaporation model for $\dot m \lesssim \dot m_{\rm crit}$}\label{sec:adaf_model}

The disk evaporation model was first proposed to explain the UV delay observed in dwarf novae \citep[][]{Meyer1994}, which was later extended in BH low-mass X-ray binaries for the spectral state transition between the low/hard state and the high/soft state, and the truncation of the accretion disk in the low/hard state \citep[][]{Meyer2000a,Meyer2000b}.

In the disk evaporation model, the electron temperature in the corona is much higher (more than four orders of magnitude) than that of the disk; a fraction of the viscously dissipated energy in the corona is transferred to the surface of the disk via electron thermal conduction. If the energy transferred to the disk surface cannot be efficiently radiated away in a thin, dense transition layer between the disk and the corona, the excessive energy will evaporate the matter in the disk continuously into the corona until an equilibrium between the disk and the corona is established (disk evaporation). Conversely, if the energy transferred to the transition layer can be efficiently radiated away, a fraction of the corona will collapse into the disk (corona condensation) \citep[][]{liuBF2002b}. The detailed region for the disk evaporation and corona condensation depends on factors such as mass accretion rate, viscosity, and magnetic field \citep[][]{Liu2007}. 
The disk evaporation model predicts an "evaporation curve", in which the evaporation rate increases with decreasing radius until a maximum evaporation rate $\dot m_{\rm crit}$ is reached. Based on the evaporation curve, if the initial accretion rate $\dot m$ is less than $\dot m_{\rm crit}$, the disk will truncate at a radius where $\dot m$ equals the evaporation rate, whereas if initial $\dot m$ is greater than $\dot m_{\rm crit}$, the disk will extend down to the ISCO of the BH with some weak corona existing above the disk \citep[][]{liuBF2002b}.

\citet[][]{taam2012} generalized the disk evaporation model from stellar-mass BHs in low-mass X-ray binaries to SMBHs in AGNs, providing general formulae for $\dot m_{\rm crit}$ and $r_{\rm tr}$, listed as follows
($r_\mathrm{tr}\equiv R/R_\mathrm{S}$ is the Schwarzschild radius-normalized radius, where $R$ is the radial distance, and $R_{\rm S}=2GM/c^2=2.95\times 10^5 M/\mathrm{M_{\odot}}\,\mathrm{cm}$ is the Schwarzschild radius, which equals to twice of the gravitational radius $R_\mathrm{g}=GM/c^2$),
\begin{equation}\label{equ:mdotcrit}
\dot m_{\rm crit} \approx 0.38\alpha^{2.34}\beta^{-0.41},
\end{equation}
\begin{equation}\label{equ:rtr}
r_{\mathrm{tr}} \approx 17.3 \dot{m}^{-0.886} \alpha^{0.07} \beta^{4.61} ,
\end{equation}
where $\alpha$ is the viscosity parameter, $\beta$ is the magnetic parameter (with magnetic pressure $p_{\rm m}={B^2/{8\pi}}=(1-\beta)p_{\rm tot}$, $p_{\rm tot}=p_{\rm gas}+p_{\rm m}$).

We calculate the truncation radius of the accretion disk $r_{\rm tr}$ with Eq.\,\ref{equ:rtr} for $\dot m \lesssim \dot m_{\rm crit}$
by specifying $\dot m$, $\alpha$, and $\beta$, the emergent spectra of the accretion flow can be calculated by combining the spectra of the inner ADAF and the outer truncated accretion disk. We adopt the self-similar solution for the structure of the ADAF, which requires input parameters $M_{\rm SMBH}$, $\dot m$, $\alpha$, $\beta$, and $\delta$. Here, $\alpha$ and $\beta$ have the same meaning and definition in the ADAF model as the disk evaporation model. 
In literature, the value of $\alpha$ ranges from a few percent to a few tenths of unity \citep{2007MNRAS.376.1740K,2019NewA...70....7M}. And for systems with weak magnetic field, $\beta$ typically adopts $0.5-1.0$ \citep[e.g.,][]{narayan&yi1995a, narayan&yi1995b, Qiao2013, 2014ARA&A..52..529Y}. $\delta$ describes the fraction of the viscously dissipated energy used to directly heat the electrons in ADAF. Since the electron mass is much smaller than the proton mass, the direct heating to the electron was initially neglected in \citet[][]{Narayan1995}, and was reincorporated in \citet[][]{mahadevan1997} and later for better matching the observations \citep[][]{Narayan1994,Narayan1995,mahadevan1997}. 
In this study, we adopt $\alpha = 0.05,\, \beta = 0.95$ and $\delta = 0.2$. These parameters are manually adjusted to improve the agreement of the predicted luminosity functions with observational data at the faint end (see Appendix.\,\ref{append:ADAF_para} for a discussion of how model parameters influence the ADAF SED).
Note that the disk evaporation model predicts a critical accretion rate $\log\dot m_\mathrm{crit} = -3.5$, based on adopted model parameters. However, to ensure continuous radiation efficiency across both slow and fast accretion regimes, we adjust $\log\dot m_\mathrm{crit}=-2.9$.

We calculate the emergent spectra of ADAF as that of in \citet[][]{Manmoto1997,Qiao2010,Qiao2013} with the method of multi-scattering of soft photons (including bremsstrahlung, synchrotron radiation of ADAF itself) by the thermal electrons in the ADAF. Appendix.\,\ref{append:sed_example} shows a few example SEDs in the low accretion regimes, compared against observational data.

\subsection{The disk-corona model for $\dot m \gtrsim \dot m_{\rm crit}$}\label{sec:disk-corona}
Above $\dot m_{\rm crit}$, a disk-corona system is formed. 
We calculate the properties of the accretion disk and the corona based on \citet[][]{LiuBF2002a,liuBF2003}.
In this model, the corona is heated by magnetic reconnection as a result of magnetorotational instability in the disk and the buoyancy of the magnetic field. Magnetic flux loops carry magnetic energy from the disk, emerging into the corona and reconnecting with each other, releasing the magnetic energy in the form of thermal energy.
This model relies on two main assumptions: 1) the heating generated by magnetic reconnection in the magnetic flux tube is cooled by Compton scattering, thermal conduction, and synchrotron radiation in the corona, and 2) before the onset of Compton cooling (i.e., when the corona has not yet fully developed), the heat is conducted downwards, heating a portion of the chromospheric plasma into the magnetic tube through a process called chromosphere evaporation (see \citet{2001ApJ...549.1160Y} for more details). Once the coronal gas density becomes sufficiently high that Compton cooling dominates over evaporation cooling, an equilibrium is established between reconnection heating and Compton cooling, after which mass evaporation at the interface ceases. In the meantime, the structure of the disk is also shaped by losing energy during magnetic reconnection. 

The \citet[][]{LiuBF2002a,liuBF2003} model exists two types of solutions, i.e., the gas-pressure dominated solution and the radiation-pressure dominated solution. For the gas-pressure dominated solution, the emission is completely dominated by the corona, and emission from the disk is very weak, nearly negligible. Gas-pressure dominated solution can exist for any accretion rate above $\dot m_{\rm crit}$, corresponding to a "hard state" with a hard X-ray spectral index $\approx 1.1$ ($\alpha_\mathrm{X}$, defined as $L_\nu\propto\nu^{-\alpha_\mathrm{X}}$) extending up to a few hundred keV. \citet[][]{liuBF2003} also found that the spectral index $\alpha_\mathrm{X}$ is nearly a constant with increasing $\dot m$ up to a few Eddington rates. However, this is inconsistent with observations, which generally suggest that $\alpha_\mathrm{X}$ increases with increasing $\dot m$, and that $\alpha_\mathrm{X}$ can be $\approx 2-3$ or even larger for $\dot m \gtrsim 0.3$ \citep{1999ApJ...526L...5L, 2004A&A...422...85P, 2004ApJ...607L.107W, 2006ApJ...646L..29S, 2008AJ....135.1505S, 2009MNRAS.399.1597S, 2011MNRAS.414.3330V, 2018MNRAS.477..210Q}.

For the radiation-pressure dominated solution, the emission is dominated by the disk, while coronal emission is nearly negligible, corresponding to a "soft state". Radiation-pressure dominated solution exists for $\dot m \gtrsim \dot{m}_\mathrm{rad}$.
$\dot{m}_\mathrm{rad}$ is determined by SMBH mass, viscosity, and magnetic field, with a value of a few tenths of the Eddington accretion rate.
The \citet{liuBF2003} model has $\dot m_\mathrm{rad} \approx 0.3$, above which the radiation-pressure dominated solution can exist at ISCO ($\approx 3R_{\rm S}$). The radiation-pressure dominated region expands outward with increasing $\dot m$. For $\dot m=1.2$, this region can extend up to $\approx 50R_{\rm S}$, beyond which the accretion flow exists in the form of the gas-pressure dominated solution.

In this study, we develop a modified mixed solution that consists of an inner radiation-pressure dominated solution plus an outer gas-pressure dominated solution for $\dot m \gtrsim \dot m_{\rm rad}$. This accretion geometry can produce a "moderate state" that matches observed X-ray spectra better. 
We update the method in \citet{liuBF2003} by solving the energy equilibria at each radial annulus instead of globally, which does not affect the accretion flow geometry.
In the radiation-pressure dominated solution, since the emission is nearly dominated by the disk and $\dot m$ has exceeded the maximum value that a standard disk can exist stably, we take the slim disk solution as a substitute. Specifically, we adopt a modified version of the self-similar solution proposed by \citep{watarai2006}, the effective temperature as a function of radius is listed as follows,
\begin{equation}\label{slimT}
T_\mathrm{eff}\approx 4.965\times10^7 \mathcal{B}^{1/8} f^{1/8} (M_\mathrm{SMBH}/\mathrm{M_\odot})^{-1/4} r^{-1/2}\mathrm{~K},
\end{equation}
where $f$ represents the fraction of the viscously dissipated energy cooled by advection. Here $\mathcal{B}=1-l_\mathrm{in}/l=1-\sqrt{R_\mathrm{in}/R}=1-\sqrt{3/r}$ is the inner boundary term, in which $l=R^2\Omega$ is the specific angular momentum, $\Omega$ is the Keplerian angular velocity, and $R_\mathrm{in}=3R_\mathrm{S}$, which differs slightly from the approach in \citet{watarai2006}, where $\mathcal{B}$ is fixed at unity. This modification reduces disk temperature at the inner radius. In addition, we correct Eq.\,20 of \citet{watarai2006} by increasing the $T_\mathrm{eff}^4$ by a factor of $\approx $2 (see Appendix.\,\ref{sec:appendSlim} for details). 

In summary, for relatively low accretion rates in the disk-corona regime, the accretion flow has the traditional disk-corona configuration as illustrated in region (2) of Fig.\,\ref{fig:geometry}.
For higher accretion rates, the accretion flow consists of an inner slim disk plus an outer disk-corona configuration, as illustrated in region (1) of Fig.\,\ref{fig:geometry}. 
In the disk-corona model, there are four parameters, i.e., $M_{\rm SMBH}$, $\dot m$, $\alpha$, and $\beta^\prime$, 
where $\beta^\prime\equiv p_\mathrm{gas}/p_\mathrm{m}$ is a magnetic parameter, defined differently from the $\beta$ parameter in Section.\,\ref{sec:adaf_model}. 
In this work, we adopt $\alpha=0.05, \beta^\prime=8$
(see Appendix.\,\ref{sec:diskcor_para} for the impact of model parameters).

Fig.\,\ref{fig:SED_diskcor} shows the SEDs of the modified magnetic reconnection-heated disk corona model for a $10^{8} M_\odot$ SMBH at accretion rates $\log{\dot{m}} = [-1.0,-0.5,0.0,1.0,2.0]$ with parameters $\alpha=0.05, \beta^\prime=8$. 
The solid curves show the total emergent spectra, while the dotted curves show the contribution from the inner slim-like region.
At relatively low accretion rates, no radiative pressure-dominated solution exists; the corresponding emergent spectrum is relatively hard, and the radiative efficiency is the same as the standard disk. 
As the accretion rate increases, the slim-like region begins to dominate, reducing the radiative efficiency and softening the spectrum.
For comparison, the hard-state SED from \citet{liuBF2003} is shown for the case $\log\dot{m}=-1.0$ with the grey dashed curve. The modified SED shows a sharper decline at high energies and a shift of the peak towards higher energies. Appendix.\,\ref{append:sed_example} shows a few example SEDs in the high accretion regimes, compared against observational data.

\subsection{Radiative efficiency and Bolometric luminosity}

\begin{figure*}
\includegraphics[width=2\columnwidth]{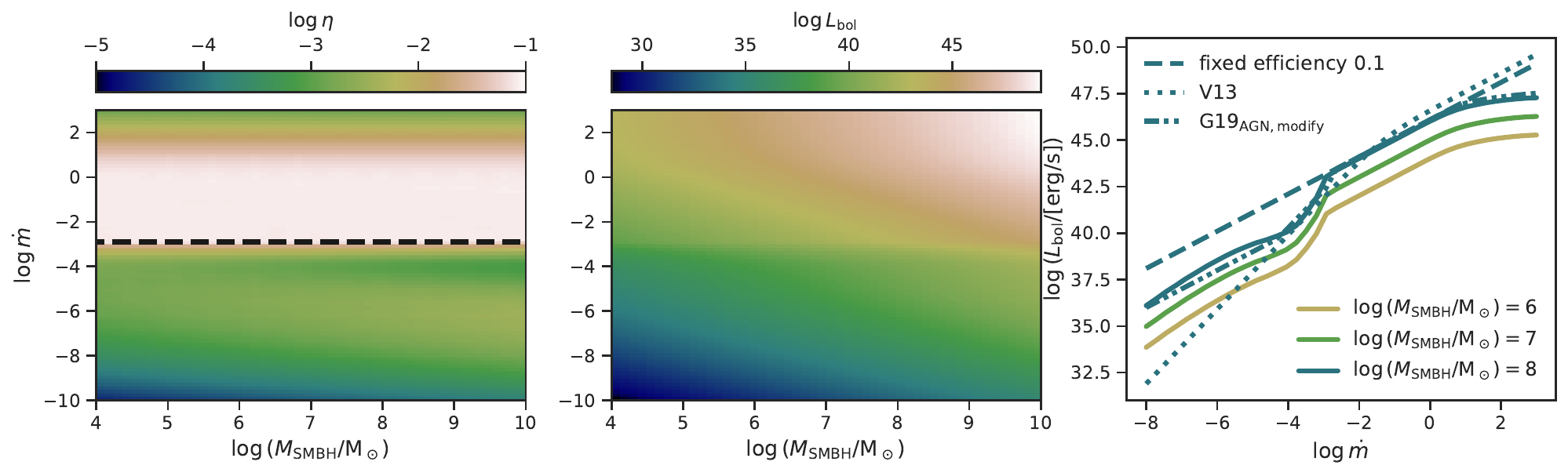}
\caption{\textit{Left:} Two-dimensional dependence of the radiative efficiency $\log\eta$ on SMBH mass and accretion rate. The black dashed line marks the boundary between the ADAF+thin disk and disk-corona regimes. The radiative efficiency shows a strong dependence on the accretion rate and a weak dependence on the SMBH mass. 
\textit{Middle:} Two-dimensional distribution of $\log L_\mathrm{bol}$ on SMBH mass and accretion rate. Both SMBH mass and accretion rate contribute comparably in determining the bolometric luminosity.
\textit{Right:} $L_\mathrm{bol}-\dot{m}_\mathrm{crit}$ relation for SMBHs with masses of $\log{(M_\mathrm{SMBH}/\mathrm{M_\odot})}=6, 7, 8$.
The solid curves represent our model predictions for SMBHs of different masses. The dashed, dotted, and dash-dotted curves depict the bolometric luminosities for an SMBH with  $\log{(M_\mathrm{SMBH}/\mathrm{M_\odot})}=8$ under various radiative efficiency models: 1) a fixed radiative efficiency $\eta=0.1$ as adopted in \citet{Henriques15, schaye2010, dobios2014, schaye2015}; 2) an accretion rate-dependent radiative efficiency as adopted in \citet{Vogelsberger2013, Vogelsberger2014} (V13); and 3) a radiative efficiency model incorporating the ADAF, standard disk, and slim disk modes, as adopted in \citet{griffin19} ($\mathrm{G19_{AGN,modify}}$), the corresponding the model parameters and critical accretion rate are identical as we adopt here.
}
\label{fig:eta_Lbol}
\end{figure*}

The radiative efficiency represents the fraction of gravitational energy transformed into radiation, which plays a key role in SMBH feedback, defined as 
\begin{equation}
    \eta \equiv L_\mathrm{bol}/\dot{M}_\mathrm{acc}c^2=\frac{L_{\rm bol}/L_{\rm Edd}}{\dot m c^2}\frac{L_{\rm Edd}}{\dot M_{\rm Edd}},
\end{equation}
where $L_\mathrm{bol}$ is the bolometric luminosity and $\dot{M}_\mathrm{acc}=\dot{m}\dot{M}_\mathrm{Edd}$ is the SMBH accretion rate in g\,s$^{-1}$. 

Previous cosmological simulations often use a simple conversion factor to convert the gravitational energy into radiation,
while our model predicts a detailed accretion rate-dependent radiative efficiency for modern cosmological galaxy formation simulations. The distribution of SMBH mass and accretion rate, color-coded by radiative efficiency, is shown in the left panel of Fig.\,\ref{fig:eta_Lbol}. The SMBH mass and accretion rate homogeneously span over $\log(M_\mathrm{SMBH}/\mathrm{M_\odot})=4-10$ and $ \log\dot{m}=-10-3$. 
In the ADAF+thin disk regime at $\log\dot{m}\le-2.9$, the radiative efficiency increases steadily with the accretion rate. In the disk-corona regime, the radiative efficiency remains around 0.1 for $-2.9<\log\dot{m}\lesssim0$, regardless of the accretion rate and SMBH mass. 
Beyond the Eddington accretion rate, the slim disk mode kicks in and plays a more prominent role. Most of the heat is directed towards the SMBH instead of being radiated away, leading to a radiation efficiency well below 0.1.

Table.\,\ref{tab:cov_eta} summarizes the Pearson correlation coefficients and corresponding p-values between radiative efficiency or bolometric luminosity and SMBH mass or accretion rate for (1) all accretion regimes, (2) the ADAF regime, (3) the disk-corona sub-Eddington regime, and (4) the disk-corona super-Eddington regime.
The radiative efficiency's dependence on SMBH mass is very weak ($r_\mathrm{mass}<0.2, p_\mathrm{mass}<10^{-4}$) in cases (1)-(3), with no statistically significant dependence ($|r_\mathrm{mass}|=2\times10^{-4},\,p_\mathrm{mass}=0.995$) in case (4); 
while the dependence of radiative efficiency on accretion rate is strong ($|r_\mathrm{macc}|>0.45,\,p_\mathrm{acc}\approx0$ in all cases).
On the other hand, the bolometric luminosity is influenced by both the accretion rate and SMBH mass ($r_\mathrm{mass/acc}\gtrsim0.4,\,p_\mathrm{mass/acc}\approx 0$) for all but the disk-corona super-Eddington regime that shows relatively smaller influence of the accretion rate ($r_\mathrm{acc}=0.2,\,p_\mathrm{acc}=10^{-9}$), reflecting the Photon-trapping effect that takes place in the central slim disk region.
The middle panel of Fig.\,\ref{fig:eta_Lbol} further demonstrates that SMBH mass and Eddington-normalized accretion rate play comparable roles in determining the bolometric luminosity of AGN. Less luminous AGNs typically have lower mass and accretion rates that are well below Eddington rates, whereas the most luminous AGNs are associated with massive SMBHs and accretion rates exceeding the Eddington level.

The right panel of Fig.\,\ref{fig:eta_Lbol} shows the bolometric luminosity increases with the accretion rate for given SMBH masses $\log(M_\mathrm{SMBH}/\mathrm{M_\odot})=6,\,7,\,8$.
The relation between the accretion rate and bolometric luminosity can be categorized into four regimes:
\begin{itemize}
    \item For super-Eddington accretion rates, $L_\mathrm{bol}$ scales logarithmically with $\dot{m}$ instead of linearly. This behavior arises from the photon trapping effect in the inner slim disk region, which occupies a significant portion of the accretion flow at such high accretion rates.
    \item At $-2.9<\log{\dot{m}}\lesssim0$, $L_\mathrm{bol}$ scales linearly with $\dot{m}$. In this accretion rate range, the accretion flow consists of a pure disk-corona system, whose radiative efficiency is explicitly controlled to be that of the standard disk (Note that the radiative efficiency depends on the spin of the SMBH, which is assumed to be zero in the scope of this paper).
    \item At $-4\lesssim\log \dot m<-2.9$, the accretion flow transitions into the ADAF regime, where the luminosity scales roughly as $L_\mathrm{bol}\propto\dot{m}^{2}$. In this range, Columb collision between ions and electrons dominates the heating term in the energy balance of electrons.
    \item At $\log{\dot{m}}\lesssim-4$, the slope flattens to $L_\mathrm{bol}\propto \dot{m}$. The heating induced by viscosity (controlled by the $\delta$ parameter) dominates over Coulomb collisions, and radiative cooling balances only the viscosity-induced heating, which roughly scales as $\propto\dot{m}$ \citep{mahadevan1997}.
\end{itemize}


For comparison, we also include the $L_\mathrm{bol}-\dot{m}$ relations adopted in the literature for an SMBH with $10^8\,\mathrm{M_\odot}$, shown in the right panel of Fig.\,\ref{fig:eta_Lbol} with different linestyles.
In \textsc{L-Galaxies}, the energy released by SMBH accretion is defined as $\eta\dot{M}_\mathrm{SMBH}c^2$, with a radiative efficiency of $\eta=0.1$, akin to that of the standard thin disk \citep{SSD1973}. This treatment is also used in hydrodynamical simulations, such as the OWLS \citep{schaye2010}, HORIZON \citep{dobios2014}, and EAGLE \citep{schaye2015} projects. The resulting $L_\mathrm{bol}-\dot{m}_\mathrm{crit}$ relation aligns with ours for moderate accretion rates $-3\lesssim\log{\dot{m}}\lesssim0$, but predicts significantly higher luminosity at lower and higher accretion rates. 
In the Illustris project \citep{Vogelsberger2013, Vogelsberger2014}, the radiative efficiency is linked to the accretion rate as $L_\mathrm{bol}=(1-\eta_\mathrm{r})\tilde{\eta}\dot{M}_\mathrm{acc}c^2$, $\tilde{\eta}=\eta_\mathrm{r} \frac{2x}{1+x}$, where $\eta_\mathrm{r}=0.2, x=\frac{1}{\chi_\mathrm{radio}}\dot{m}, \chi_\mathrm{radio}=0.05$. The expected relation matches ours for $-4\lesssim\log{\dot{m}}\lesssim0$, but diverges significantly at higher and lower accretion rates.
\citet{griffin19} ($\mathrm{G19_{AGN,modify}}$) introduced a comprehensive model that includes ADAF, standard disk, and slim disk models, displaying a consistent $L_\mathrm{bol}-\dot{m}$ relation with our results.

\begin{table*}
\begin{center}
\caption{Correlation coefficients between radiative efficiency $\log\eta$/bolometric luminosity $\log L_\mathrm{bol}$ and SMBH mass $\log(M_\mathrm{SMBH}/\mathrm{M_\odot})$/accretion rate $\log\dot{m}$.}
\label{tab:cov_eta}
\begin{tabular}{lcccccccc}
\hline
\hline
 & \multicolumn{4}{c}{Radiative efficiency} & \multicolumn{4}{c}{Bolometric luminosity} \\
 & $r_\mathrm{mass}$ & $r_\mathrm{acc}$ & p-value$_\mathrm{mass}$ & p-value$_\mathrm{acc}$ & $r_\mathrm{mass}$ & $r_\mathrm{acc}$ & p-value$_\mathrm{mass}$ & p-value$_\mathrm{acc}$\\
\hline
\hline
All regimes & 0.06 & 0.75 & $10^{-4}$ & 0.0 & 0.37 & 0.92 & 0.0 & 0.0\\
\hline
ADAF & 0.18 & 0.81 & $9\times10^{-17}$ & 0.0 & 0.59 & 0.80 & 0.0 & 0.0\\
\hline
diskcor\_subEdd & 0.16 & 0.45 & $2\times10^{-6}$ & $6\times10^{-44}$ & 0.91 & 0.42 & 0.0 & $8\times10^{-37}$\\
\hline
diskcor\_supEdd & $-2\times10^{-4}$ & -0.98 & 0.995 & 0.0 & 0.98 & 0.20 & 0.0 & $10^{-9}$\\
\hline
\hline
\end{tabular}
\end{center}
\end{table*}


\section{results} \label{sec:lum}
To enable a proper comparison between models and observations, it is necessary to convert the model-predicted physical properties - SMBH mass and accretion rate - into observables. 
We apply the AGN SED model in post-processing, utilizing black hole masses and accretion rates from the semi-analytic galaxy catalog to study AGN bolometric corrections, detection rates, and luminosity functions, and compared them with observations.

{
\subsection{Optical-X-ray Spectra Index}
\begin{figure}[h]
\centering
\includegraphics[width=\columnwidth]{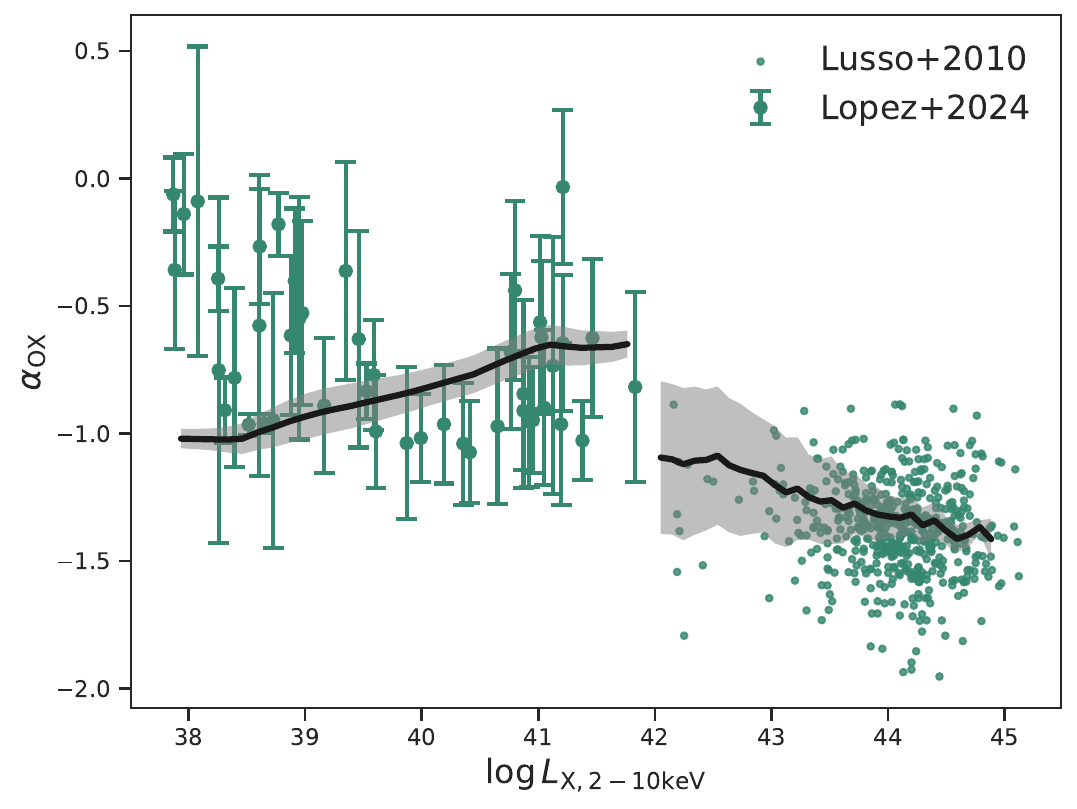}
\caption{\textbf{relation between 2-10\,keV X-ray luminosities and the spectra indices between 2500\AA-2\,keV of LLAGNs and HLAGNs.} 
The black solid curve and the grey contour region denotes the median and $\pm1\sigma$ values of the $\log{L_\mathrm{X, 2-10keV}}-\alpha_\mathrm{OX}$ distribution predicted by the ADAF model and disk-corona model at $z\approx0$, overplotted with the best-fitted results from a group of local LLAGN derived by \citet{2024A&A...692A.209L} and HLAGN ($L_\mathrm{2-10keV}>10^{42}\mathrm{erg\,s^{-1}}$) at $0.4\lesssim z\lesssim2$ by \citet{2010A&A...512A..34L}. The model prediction is computed with AGN populations that have similar X-ray luminosity, SMBH mass, and accretion rate ranges as the observational sample in \citet{2010A&A...512A..34L, 2024A&A...692A.209L}.
\label{fig:alpha_OX}}
\end{figure}

The spectral index characterizes how the radiative flux or luminosity varies with frequency within a given wavelength band, i.e., $F_\nu\propto \nu^{\alpha};\ L_\nu\propto\nu^\alpha$, which is determined by the AGN SED. It is pivotal in many observational studies, such as in performing K-corrections to observed luminosities and deriving bolometric luminosities using bolometric corrections.
A specific case is the optical-X-ray spectral index, $\alpha_\mathrm{OX}$, which refers to the slope between the X-ray luminosity at 2 keV and the optical luminosity at 2500\,\AA. This index provides insights into the relationship between the optical and X-ray emissions of AGNs, reflecting the efficiency of the emission mechanisms across different wavelengths.
In this section, we compare the $\alpha_\mathrm{OX}$ predicted by our model with those determined observationally. This comparison serves as a critical test of the model’s accuracy in reflecting the observed AGN behavior.

Fig.\,\ref{fig:alpha_OX} presents the $\log L_\mathrm{X, 2-10,keV}$–$\alpha_\mathrm{OX}$ relation predicted by our model, where the black solid curve denotes the median values and the grey shaded region indicates the $\pm1\sigma$ scatter. Observational data are overplotted as scatter points, including a local LLAGN sample from \citet{2024A&A...692A.209L}, derived via SED fitting, and a HLAGN sample ($L_\mathrm{2-10keV}>10^{42}\mathrm{erg\,s^{-1}}$) at $0.3 \lesssim z \lesssim 2$ from \citet{2010A&A...512A..34L}, based on multiwavelength SEDs. For a fair comparison with the LLAGNs, we further restrict the model AGN sample to a similar SMBH mass range of $7 < \log(M_\mathrm{SMBH}/M_\odot) < 9$, and limit our analysis to SMBHs in the ADAF regime, consistent with the AGN properties suggested by \citet{2024A&A...692A.209L}.


In both the low- and high-luminosity regimes, our model predictions are generally consistent with the observational results, although the observed $\alpha_\mathrm{OX}$ values tend to be slightly higher than the predicted ones at the faintest end. We note that the $\alpha_\mathrm{OX}$-luminosity relation, particularly at the faint end, may be affected by selection effects and other observational limitations. In summary, our model reproduces the $\alpha_\mathrm{OX}$-luminosity relation well for both LLAGNs and HLAGNs, achieving good agreement across several orders of magnitude in luminosity.

\subsection{Bolometric corrections}
\begin{figure*}
\includegraphics[width=2\columnwidth]{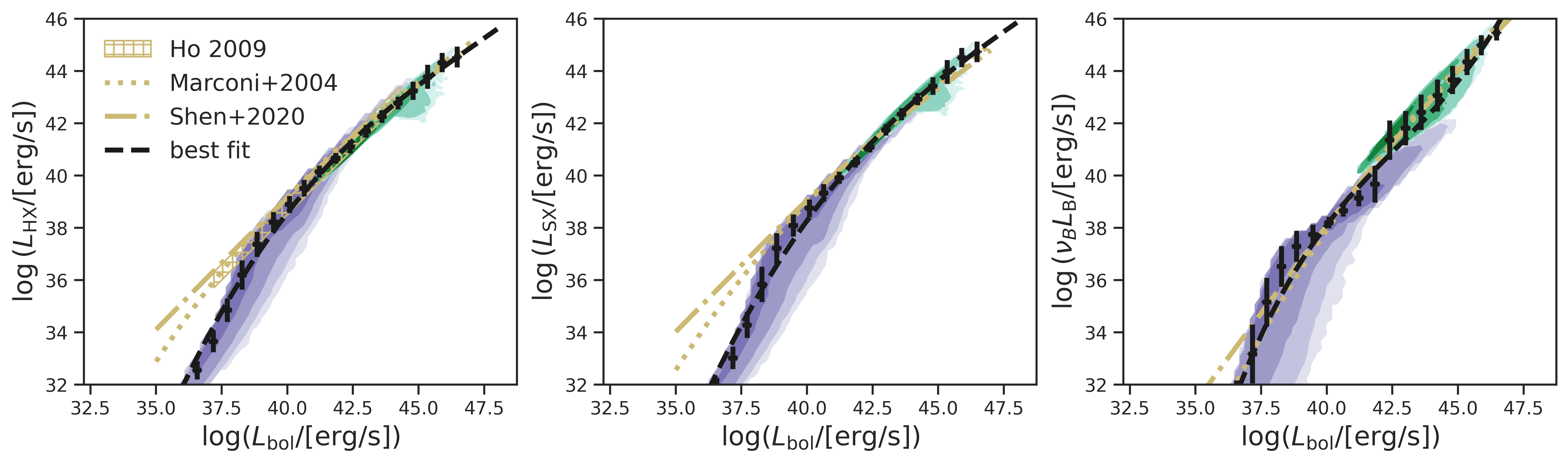}
\caption{\textbf{Relations between bolometric luminosities and luminosities at different wavelengths predicted by our combined galaxy formation + AGN SED model} (left panel: bolometric - hard X-ray; middle panel: bolometric - soft X-ray; right panel: bolometric - B band). 
The green and purple contour regions represent the distribution of bolometric luminosity-photometric luminosity pairs calculated from the $z=0$ snapshot, for the disk-corona and ADAF regimes respectively. The black curves represent the best-fitted bolometric corrections.
For comparison, bolometric corrections from \citet{marconi2004} and \citet{2020MNRAS.495.3252S} are shown as yellow dotted and dash-dotted curves respectively. The hatched region in the left shows the bolometric corrections adopted in \citet{2009ApJ...699..626H}, which is derived from observed broad-band LLAGN SEDs.}
\label{fig:scaling_relation}
\end{figure*}

Observational works often require a bolometric correction factor to derive the AGN's bolometric luminosity for constructing the bolometric luminosity function and accretion rate distribution \citep[e.g.,][]{2017ApJ...846..112K,2020ApJ...900..124B, marconi2004, hopkins2007, 2020MNRAS.495.3252S}. These corrections are typically calculated based on an SED model with observed spectra slope \citep[e.g.,][]{2001AJ....122..549V, Krawczyk2013, Lusso2015}.
On the other hand, theoretical works often use bolometric correction factor to convert bolometric luminosity from SMBH properties to photometries at specific bands for direct comparison with observations \citep[e.g.,][]{hirschmann2014, griffin19}.
It is crucial to acknowledge the existence of different SEDs for the same bolometric luminosity (see Appendix.\,\ref{sec:bols}), which leads to potential dispersion in relations between bolometric luminosity and luminosities across various wavelength bands.

By combining the \textsc{L-Galaxies} galaxy catalog and our AGN SED model,
Fig.\,\ref{fig:scaling_relation} shows the predicted relations between bolometric luminosity and photometry in specific bands: hard X-ray (2-10 keV), soft X-ray (0.5-2 keV), and B band (4400 \AA) from left to right. Bright AGNs occupy the disk-corona regime, while less luminous AGNs are found in the ADAF+thin disk regime. In the middle parts, both modes contribute, especially to the hard X-ray and soft X-ray bands. 

The green and purple contour regions represent the distribution of bolometric luminosity-photometric luminosity pairs calculated from the $z=0$ snapshot, for the disk-corona and ADAF regimes respectively. 
The regions enclosed by the deepest contour regions correspond to histogram counts of $\approx50000$ and $\approx5000$ for ADAF and disk-corona respectively.
The black crosses indicate median values, with errorbars representing twice the standard deviations. Significant scatter exists at both high and low luminous ends, up to an order of magnitude. This issue is often overlooked in past research. We fit the bolometric corrections with a third-degree polynomial function, denoted by the black dashed curves.
The fitting process employs the Markov Chain Monte Carlo (MCMC) method \citep{Foreman_Mackey_2013} to derive the best-fit parameters. The posterior distribution in the MCMC process is assumed to follow a Gaussian distribution, characterized by the median value and the standard deviations of the data in each bin.
\begin{equation}\label{eq:scaling}
\begin{aligned}
\log\left(L_{\mathrm{HX}} / L_{\mathrm{bol}}\right)=& - 1.700^{+0.166}_{-0.165} - 0.264^{+0.136}_{-0.136} \mathcal{L}\\
& -0.018^{+0.035}_{-0.035} \mathcal{L}^2 + 0.004^{+0.002}_{-0.002} \mathcal{L}^3,\\
\log\left(L_{\mathrm{SX}} / L_{\mathrm{bol}}\right)=& - 1.497^{+0.164}_{-0.164} - 0.215^{+0.133}_{-0.133} \mathcal{L}\\
& - 0.030^{+0.035}_{-0.035} \mathcal{L}^2 + 0.003^{+0.002}_{-0.002} \mathcal{L}^3 \\
\log\left(\nu_\mathrm{B} L_{\nu_\mathrm{B}} / L_{\mathrm{bol}}\right)=& - 1.212^{+0.178}_{-0.178} + 0.388^{+0.175}_{-0.175} \mathcal{L}\\
& + 0.130^{+0.053}_{-0.053} \mathcal{L}^2 + 0.015^{+0.004}_{-0.004} \mathcal{L}^3,
\end{aligned}
\end{equation}
where $\mathcal{L}=\log{(L_\mathrm{bol}/10^{12}\mathrm{L}_\odot)}$ is the bolometric luminosity, $L_\mathrm{HX}$ is the hard X-ray (2-10 keV) luminosity, $L_\mathrm{SX}$ is the soft X-ray (0.5-2 keV) luminosity, and $L_{\nu_\mathrm{B}}$ is the luminosity per unit frequency centered at $\nu_\mathrm{B}=c/4400$\AA. The coefficient uncertainties represent the variation in the bolometric corrections.

We also include the bolometric corrections in previous literature for comparison \citep{marconi2004, 2020MNRAS.495.3252S}. In \citet{marconi2004}, the bolometric corrections are derived using a template AGN SED that consists of a broken power-law in the optical-UV region $L_\nu\propto\nu^\alpha$ ($\alpha=-0.44$ for $1\mu\mathrm{m}<\lambda<1300\text{\AA}$; $\alpha=-1.76$ for $500\text{\AA}<\lambda<1200\text{\AA}$), a truncated tail ($\alpha=2$ for $\lambda>1\mu\mathrm{m}$), and X-ray spectrum ($>1$keV) that consist of a single power law ($\Gamma=1.9$) and an exponential cutoff at $E_c=500\mathrm{keV}$. It aligns with our median luminosity values for bright AGNs but overestimates luminosity in both hard and soft X-ray bands for faint AGNs. The SED model in \citet{2020MNRAS.495.3252S} combines spectra templates from various photometric bands. Optical/UV and IR templates are from \citet{Krawczyk2013} and \citet{2006ApJS..166..470R}, while the X-ray component includes a cut-off power-law model with $\Gamma=1.9$ and $E_\mathrm{c}=300\,\mathrm{keV}$ \citep{2008A&A...485..417D, ueda2014, aird2015}, along with a reflection component using the \textsc{pexrav} model \citep{1995MNRAS.273..837M}, all connected to the Optical/UV band. Like \citet{marconi2004}, it aligns with our median luminosity values for bright AGNs but predicts greater luminosity in both hard and soft X-ray bands for faint AGNs. Our B-band luminosity results align with previous studies by \citet{marconi2004} and \citet{2020MNRAS.495.3252S} at all luminosities. 
The hatched region in the left shows the hard X-ray bolometric correction adopted in \citet{2009ApJ...699..626H}, which is derived from observed broad-band LLAGN SEDs. The X-ray luminosity scales linearly with bolometric luminosity as $C_\mathrm{HX}=L_\mathrm{bol}/L_\mathrm{HX}=15.8$, with 0.3 dex uncertainties, also aligns with our results in their sample luminosity range $L_\mathrm{bol}\approx 10^{37}-3\times10^{44}\mathrm{erg\,s^{-1}}$.
In sum, our results align with the literature at a 1$\sigma$ level for bolometric luminosities exceeding $10^{40}\mathrm{erg\,s^{-1}}$ across all bands, where observational data provide stronger constraints. Our model predicts lower X-ray luminosities at low accretion rates than previous studies.

\subsection{Bright AGN fraction in massive galaxies}\label{sec:f_agn}

\begin{figure}
\includegraphics[width=\columnwidth]{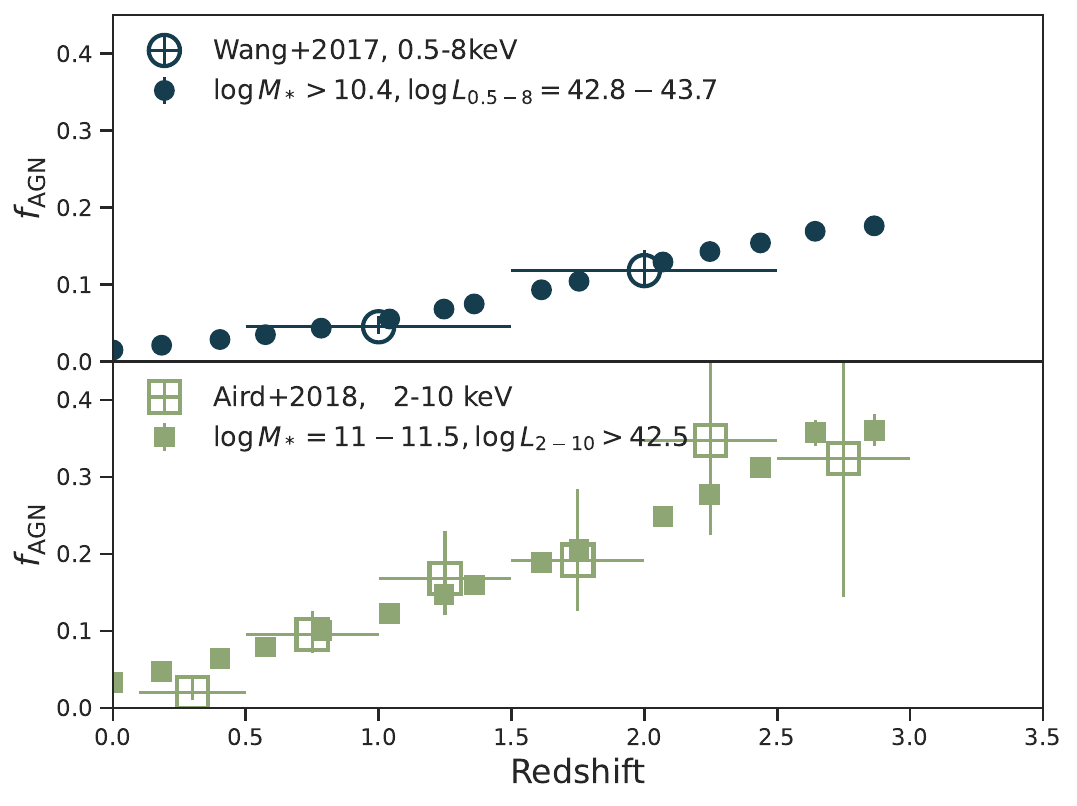}
\caption{\textbf{Fraction of bright AGN in massive galaxies.}
\textit{Upper:} The blue open symbols show the AGN fraction collected from \citet{2017A&A...601A..63W}, and the blue filled symbols show the model prediction using the same selection criterion $M_*>10^{10.6}\,\mathrm{M_\odot},\ L_\mathrm{0.5-8}=10^{42.8}-10^{43.7}\,\mathrm{erg\,s^{-1}}$; 
\textit{Lower:} The green open symbols shows the AGN fraction derived using the probability distribution function of specific BH accretion rate in \citep{2018MNRAS.474.1225A}, and the blue filled symbols show the model prediction using the same selection criterion $M_*=10^{11}-10^{11.5}\,\mathrm{M_\odot},\ L_\mathrm{2-10}>10^{42.5}\,\mathrm{erg\,s^{-1}}$. The errorbars indicate the 1$\sigma$ Poisson uncertainties.
\label{fig:f_AGN}}
\end{figure}

Detailed SED modeling facilitates an in-depth examination of the active AGN fraction, allowing for direct comparison with observations. 
Fig.\,\ref{fig:f_AGN} presents the fractions of luminous AGN in massive galaxies predicted by our model alongside observational measurements obtained at (slightly) different wavelength ranges. We intentionally do not homogenize these data to a single wavelength band, as our model provides the full AGN SED and can therefore be compared directly to observations in their native bands.

The upper panel of Fig.\,\ref{fig:f_AGN} shows the fraction of AGN with X-ray luminosities $L_{\mathrm{0.5-8\,keV}} = 10^{42.8}-10^{43.7}\,\mathrm{erg\,s^{-1}}$ hosted by galaxies with stellar masses $M_\ast > 10^{10.6}\,M_\odot$ at $z = 0.5-1.5$ and $z=1.5-2.5$. 
The observational data are collected directly from \citet{2017A&A...601A..63W}, who applied detailed corrections for various selection effects using the deep X-ray and UV-to-far-infrared data in the two Great Observatories Origins Deep Survey (GOODS) fields. 

The bottom panel compares the fraction of luminous AGN with $L_{\mathrm{2-10\,keV}} > 10^{42.5}\,\mathrm{erg\,s^{-1}}$ hosted by galaxies with stellar masses $M_\ast = 10^{11}-10^{11.5}\,M_\odot$. 
 The fraction of luminous AGN is computed by integrating the probability distribution function (PDF) of the 
specific black hole accretion rate, $\lambda_{\mathrm{sBHAR}}$, from \citet{2018MNRAS.474.1225A}, which was 
derived from deep \textit{Chandra} observations of stellar mass-selected galaxies in the CANDELS and 
UltraVISTA surveys. The specific accretion rate $\lambda_{\mathrm{sBHAR}}$ is defined in terms of the 
2-10\,keV X-ray luminosity $L_{\mathrm{2-10\,keV}}$ and the host galaxy stellar mass $M_*$. Integrating the PDF therefore 
allows us to estimate the AGN fraction above a given $L_{\mathrm{2-10\,keV}}$ threshold for any fixed $M_*$ as,
\begin{equation}
f_{\mathrm{AGN}}(>L_{\mathrm{X}} \mid M_\ast,z)
= \int_{\lambda_{\mathrm{sBHAR,X}}}^{\infty}
p(\lambda_{\mathrm{sBHAR}} \mid M_\ast, z)\, d\lambda_{\mathrm{sBHAR}} \, ,
\end{equation}

Across both panels, our model predictions show good agreement with the observational results. This consistency demonstrates the predictive ability of our SED-based framework, highlighting its ability to reproduce the observed fraction of luminous AGN in massive galaxies over a broad range of wavelengths.

} 

\subsection{Luminosity functions}\label{sec:LF}


\begin{figure*}
\includegraphics[width=2\columnwidth]{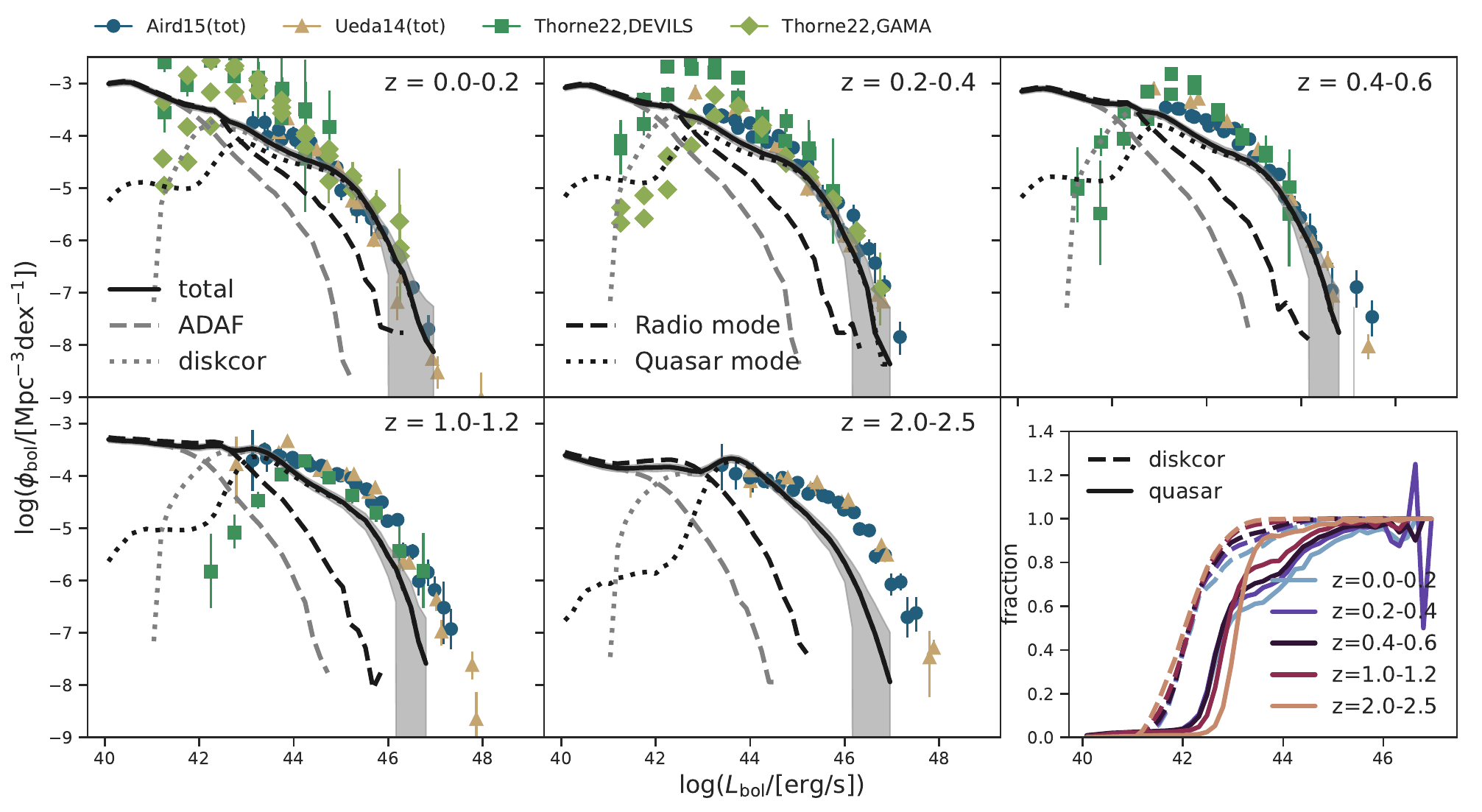}
\caption{{\bf Bolometric luminosity functions at different redshifts.}  
Model predictions of total luminosity functions are shown in the first five panels with solid black curves. Contributions from radio-mode and quasar-mode accretion are depicted with black dashed and dotted curves, while the contributions from ADAF+thin disk and disk-corona models are represented by grey dashed and dotted curves. The grey shaded area accounts for the predicted Poisson error and cosmic variance.
The final panel illustrates the fraction of SMBH abundance in the disk-corona regime (dash curves) and quasar-mode (solid curves) as a function of bolometric luminosity across various redshift intervals. 
The scatter points show the observed bolometric luminosity functions that are converted from the X-ray luminosity functions of all AGNs in \citet{ueda2014, aird2015} (see Section.\ref{sec:XLF}), using the bolometric correction in \citep{marconi2004}; and the bolometric luminosity functions in \citet{Thorne2022}, estimated by fitting the FUV to FIR spectra.
\label{fig:LF_bol}}
\end{figure*}

\begin{figure*}
\includegraphics[width=2\columnwidth]{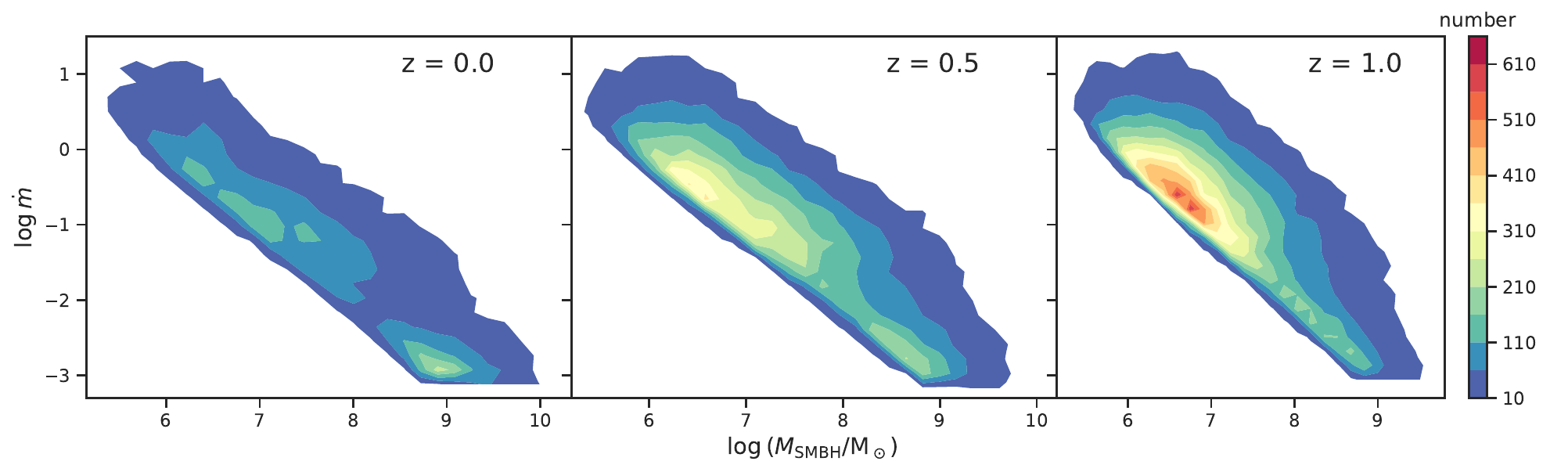}
\caption{{\bf Joint distribution of SMBH mass and accretion rate for luminous AGNs}. It shows the results for luminous AGNs with $\log{(L_\mathrm{bol}/\mathrm{[erg\,s^{-1}]})}>44$ at redshift $z=0$ (left), $z=0.5$ (middle), and $z=1$ (right). The contour curves are color-coded according to the SMBH abundance.
\label{fig:Lbol_z}}
\end{figure*}

\begin{figure*}
\includegraphics[width=2\columnwidth]{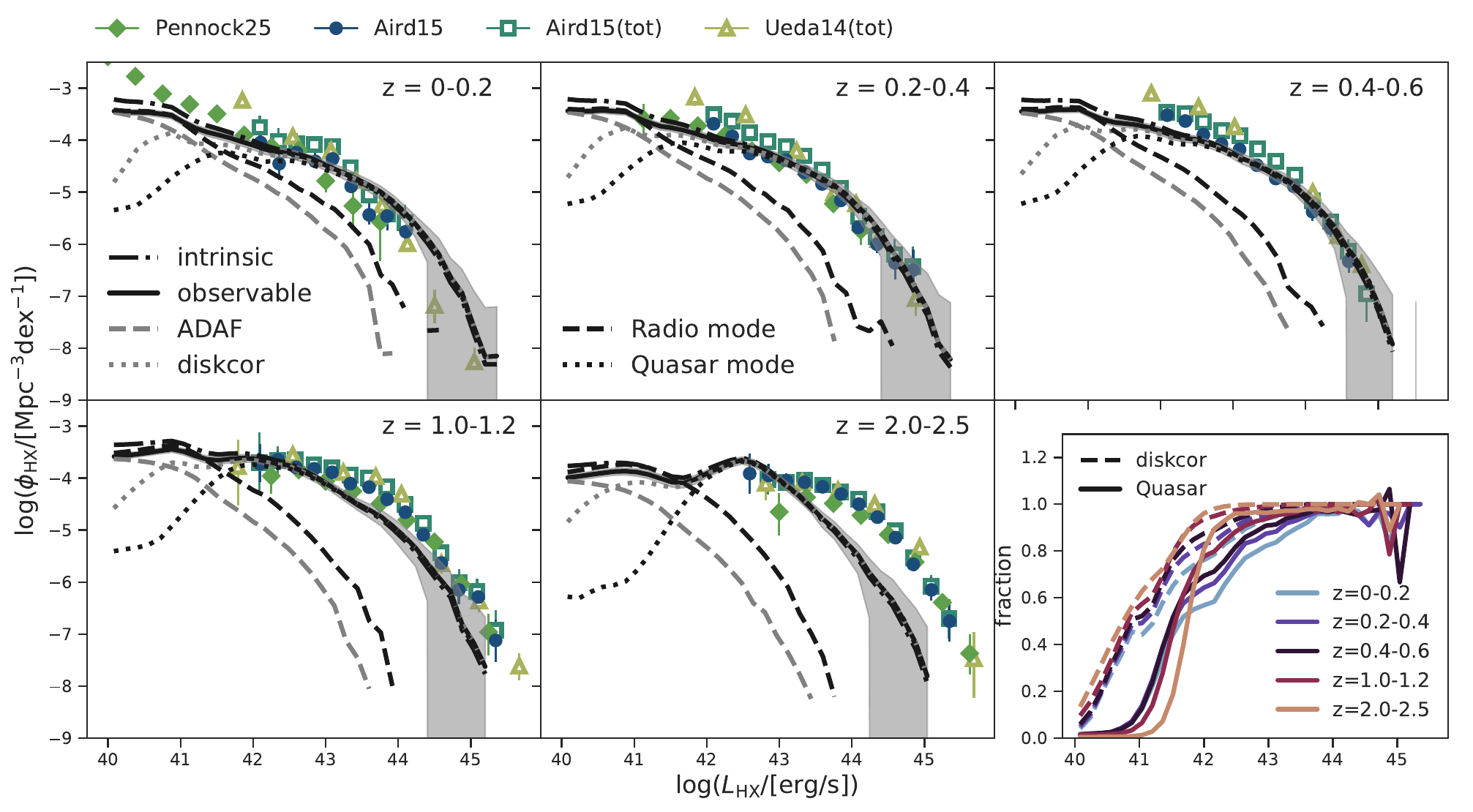}
\caption{{\bf Rest-frame hard X-ray [2-10 keV] luminosity functions at different redshifts.}  
Similar to the bolometric luminosity functions in Fig.\,\ref{fig:LF_bol}, but for the hard X-ray luminosities. Black solid curves present the results adopting the visible fraction from \citet{hopkins2007}. The intrinsic luminosity functions of all AGNs are denoted with black dash-dotted curves.
Filled scatters show the observational data that have not been corrected for absorption effects, taken from \citet{aird2015, 2025MNRAS.tmp.1734P}; the open scatters are taken from \citet{ueda2014, aird2015}, showing the binned estimations of all AGNs.
\label{fig:LF_HX}}
\end{figure*} 

\begin{figure*}
\includegraphics[width=2\columnwidth]{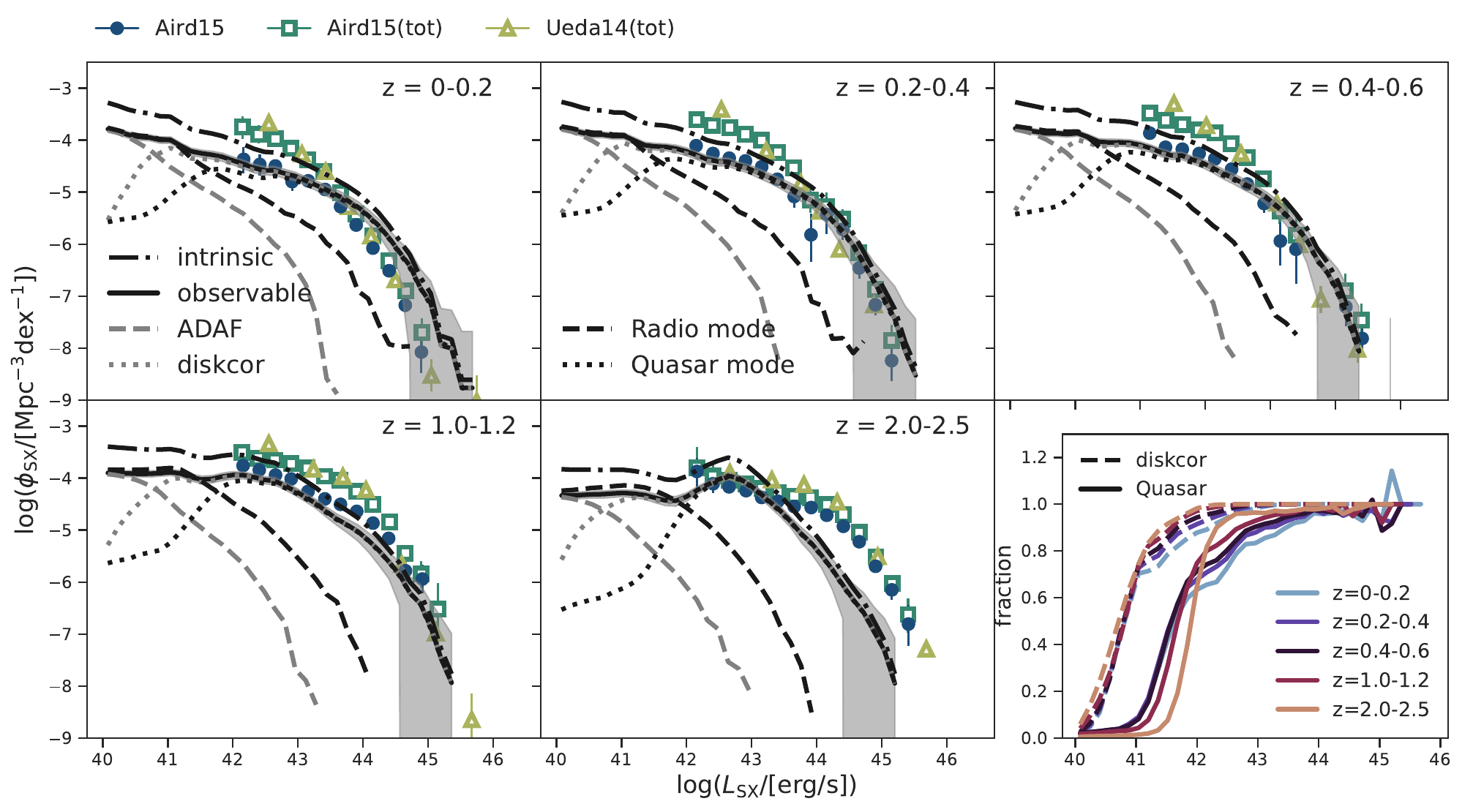}
\caption{{\bf Rest-frame soft X-ray [0.5-2 keV] luminosity functions at different redshifts.}  
Similar to the bolometric luminosity functions in Fig.\,\ref{fig:LF_bol}, but for the soft X-ray luminosities. Black solid curves present the results adopting the visible fraction from \citet{hopkins2007}. The intrinsic luminosity functions of all AGNs are denoted with black dash-dotted curves.
Filled scatters show the observational data that have not been corrected for absorption effects, taken from \citet{hasinger2005, ueda2014}, and the open scatters are taken from \citet{ueda2014, aird2015}, showing the binned estimations of all AGNs.
\label{fig:LF_SX}}
\end{figure*}

SED for individual AGN allows us to directly determine bolometric, X-ray, and optical luminosities by integrating over corresponding wavelengths rather than relying on bolometric corrections. Here, we analyze the predicted luminosity functions and compare them with observations while investigating the impacts of different accretion modes. 
In practice, at each redshift interval of interest, the AGN luminosity functions are constructed by binning the sources into intervals of 0.16 dex in luminosity for the bolometric and X-ray bands, and 0.27 magnitudes for the optical band. The binned counts are subsequently normalized by the comoving volume.

\subsubsection{AGN obscuration and visible fractions}\label{sec:obscuration}

AGNs are believed to be surrounded by a dusty torus that absorbs radiation along certain sightlines. For simplicity, we assume that, at a given wavelength, AGNs are either fully obscured or fully visible. Thus, obscuration can be described in terms of a visible fraction, defined as the proportion of sources that remain unobscured in a given band at a specified luminosity and redshift. 

In practice, we adopt the visible fractions derived by \citet{hopkins2007},
\begin{equation}
    f(L_\mathrm{bol}) = f_{46}\left(\frac{L_\mathrm{bol}}{10^{46}\,\mathrm{erg\,s^{-1}}}\right)^\beta
\end{equation}
with $(f_{46}, \beta) = (0.609, 0.063)$ for the soft X-ray band and $(f_{46}, \beta) = (1.243, 0.066)$ for the hard X-ray band. For the 1500\AA~band, we adopt $(f_{46}, \beta) = (0.15, -0.1)$ as suggested by \citet{griffin19}, who calibrated parameters of the visible fraction to ensure consistency between the 1500{\AA}-derived bolometric luminosity function and those obtained from other bands. The observed luminosity functions are then obtained from the total AGN LFs through  
\begin{equation}
\phi_\mathrm{obs} = f(L_\mathrm{bol}) \times \phi_\mathrm{total},
\end{equation} 
 where $\phi_\mathrm{tot}$ is the intrinsic AGN LF predicted by the model that consists of all AGN populations, and $\phi_\mathrm{obs}$ refers to the LF after accounting for obscuration effects.

\subsubsection{Bolometric luminosity function}

We present the bolometric luminosity functions predicted by our model and compare them with the observational estimates in Fig.\,\ref{fig:LF_bol}. 
Model-predicted bolometric luminosities are calculated by directly integrating AGN SEDs.
Error estimates (grey shaded areas) include Poisson error and cosmic variance  $\sigma=\sqrt{\sigma_\mathrm{CV}^2+\sigma_\mathrm{P}^2}$. The cosmic variance is calculated by splitting the simulation box into 125 sub-volumes of approximately $96\mathrm{Mpc/h}$ on each side.
The observational data are converted from hard and soft X-ray observations \citep{ueda2014, aird2015}, utilizing the bolometric corrections derived in \citet{marconi2004}.
We also take the bolometric luminosity functions estimated by \citet{Thorne2022}, which are converted from the FUV to FIR observation using the SED fitting code \textsc{ProSpect}.

Our model successfully reproduces the bolometric luminosity function up to $z\approx1.2$, but struggles to predict the number density accurately for $z>2$, particularly at high luminosities. 

The observation data primarily focus on luminous AGNs. The contour plot in Fig.\,\ref{fig:Lbol_z} illustrates the distribution of SMBH mass and accretion rate for AGNs with $\log(L_\mathrm{bol} / \mathrm{[erg\,s^{-1}]}) > 44$. The number of luminous AGNs increases with redshift between 0 and 1.5. Luminous AGNs at $z>0.5$ are mainly in SMBHs with $\log(M_{\rm SMBH}/\mathrm{M_{\odot}})\approx 6-7$ and near the Eddington accretion rates. At $z=0$, both high-mass SMBHs with low accretion rates and intermediate-mass SMBHs of $\log(M_{\rm SMBH}/\mathrm{M_{\odot}}) \approx 6-7$ with high accretion rates contribute to bright AGNs.


We further distinguish contributions to bolometric luminosities based on physical accretion mechanisms: radio-mode vs. quasar-mode accretion. Contributions are also categorized by the ADAF+thin disk model and the disk-corona model. Accretion rates in the radio-mode are usually low and are primarily associated with ADAF+thin disk accretion. In contrast, the quasar-mode has higher accretion rates, aligning with the disk-corona mode. It is evident that radio-mode accretion/ADAF+thin disks dominate the faint end ($\log{(L_\mathrm{bol}/\mathrm{[erg\,s^{-1}]})}\lesssim 42$) of the bolometric luminosity functions across all redshifts. On the other hand, the quasar-mode accretion/disk-corona model dominates in the intermediate to bright part of the luminosity functions. 

More quantitative comparisons are presented in the last panel of Fig.\,\ref{fig:LF_bol}. The contribution from disk-corona model accretion increases rapidly with luminosity below $\log(L_\mathrm{bol}/\mathrm{[erg\,s^{-1}]})\approx43$, becoming flatter at higher luminosities and reaching 100\% above $\log(L_\mathrm{bol}/\mathrm{[erg\,s^{-1}]})\approx44$. In the range of $\log(L_\mathrm{bol}/\mathrm{[erg\,s^{-1}]})\approx42-44$, its contribution exhibits a positive redshift dependency.

Interestingly, below $\log(L_\mathrm{bol}/\mathrm{[erg\,s^{-1}]})\approx44$, the contribution from disk-corona mode exceeds that from quasar-mode accretion, indicating high rates of radio-mode accretion that activate the disk-corona mode of AGN activity. The saturation luminosity (the luminosity at which the corresponding contribution from disk-corona/quasar-mode reaches $\approx100\%$ in the last panel of Fig.\,\ref{fig:LF_bol}) is higher in quasar-mode accretion (log$(L_\mathrm{bol}/\mathrm{[erg\,s^{-1}]})\approx46$) compared to disk-corona mode. A stronger redshift dependence is also observed in quasar-mode accretion than in disk-corona mode between $\log(L_\mathrm{bol}/\mathrm{[erg\,s^{-1}]})\approx41.5-44$.

\subsubsection{X-ray luminosity functions}\label{sec:XLF}
Previous cosmological simulations often rely on bolometric corrections to convert bolometric luminosity to X-ray luminosities. In this work, we directly calculate the X-ray emission by integrating the AGN SED. 

Fig.\,\ref{fig:LF_HX} shows our model's predicted rest-frame hard X-ray (2-10 keV) luminosity functions at different redshifts.  Black solid curves denote the modeled $\phi_\mathrm{obs}$, which are compared with the observed results from \citet{aird2015, 2025MNRAS.tmp.1734P}. The black dash-dotted curves represent $\phi_\mathrm{total}$, and are compared with the observational results of \citet{ueda2014, aird2015} for all AGNs.

Our model reproduces the observed hard X-ray luminosity functions up to redshifts $z \lesssim 1$, but underestimates the abundance of luminous AGN (i.e., $L_{\mathrm{2-10\,keV}} > 10^{42.5}\,\mathrm{erg\,s^{-1}}$)  at higher redshifts $z \gtrsim 1$. At first glance, this may appear inconsistent with the good agreement shown in the bottom panel of Fig.\,\ref{fig:f_AGN} between the predicted luminous AGN fractions and the corresponding observational measurements. However, we note that AGN fractions are not strongly influenced by the brightest AGNs that dominate the discrepancy in the X-ray luminosity functions. The number of AGN just above the luminosity thresholds used in the observational analyses (i.e.\ $L_{\mathrm{X}} \sim 10^{42.5}-10^{43}\,\mathrm{erg\,s^{-1}}$) is much larger than that of those with higher luminosities. As a result, AGN fractions are primarily sensitive to this more numerous, moderately luminous population, where our model and the observations are in much better agreement. The consistency between the predicted and observed AGN fractions is therefore fully compatible with the residual tension at the bright end of the X-ray luminosity function for $z>1$.
This aligns with the comparison of bolometric luminosity functions, with a more significant deviation from observations at higher redshifts. The contribution from quasar/disk-corona mode accretion follows a similar trend to bolometric luminosities, with a slower increase before reaching saturation.
Likewise, Fig.\,\ref{fig:LF_SX} shows the rest-frame soft X-ray luminosity functions at different redshift intervals with the same line conventions, where the predicted $\phi_\mathrm{obs}$ is compared with observations from \citet{ueda2014, aird2015} that are not corrected for obscuration, and $\phi_\mathrm{tot}$ is compared with observations from \citet{ueda2014, aird2015} for all AGNs.
Similar to the hard X-ray luminosity functions, our model aligns well with observed soft X-ray luminosity functions up to $z = 1$ but falls short at higher redshifts. The contribution from quasar/disk-corona mode accretion increases more slowly with soft X-ray luminosity compared to bolometrics, but faster than in hard X-ray before reaching saturation.

\subsubsection{1500{\AA} luminosity function}

\begin{figure*}
\includegraphics[width=2\columnwidth]{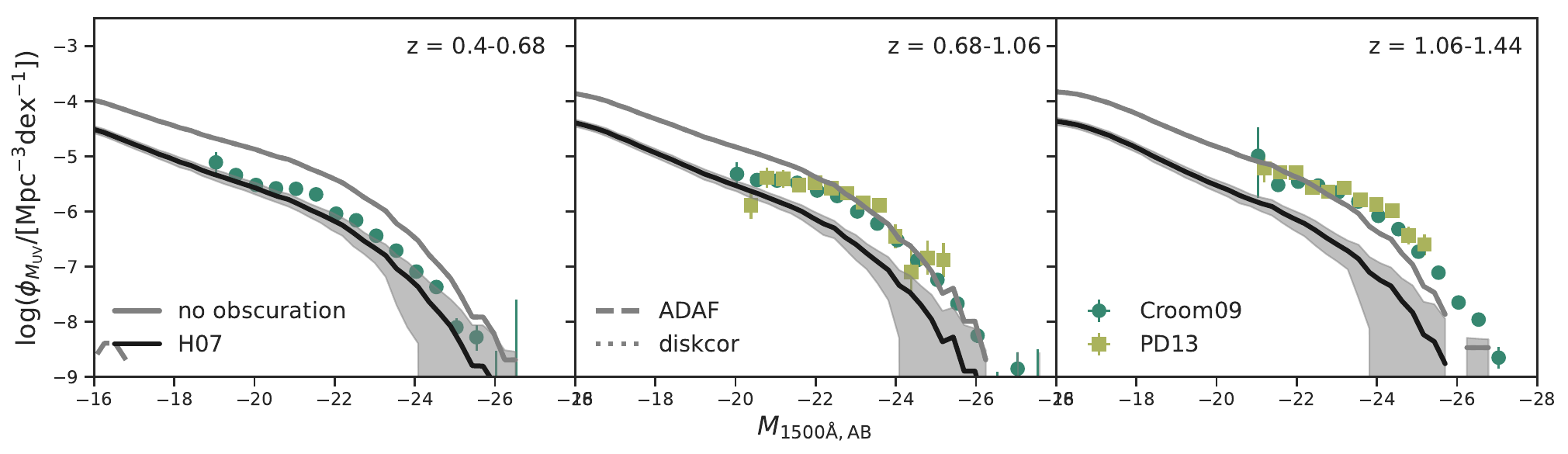}
\caption{{\bf Rest-frame 1500{\AA} luminosity functions at different redshifts.}
The grey solid curves display the total luminosity function, with different accretion models denoted by varied linestyles (ADAF+thin disk - dashed; disk-corona model - dotted, overlapped with the grey solid curve). The black solid curves show the luminosity functions adopting the Hopkins-style visible fraction in \citet{griffin19}.
Observation data are taken from \citet{croom2009, 2013A&A...551A..29P} without correction for AGN obscuration.
\label{fig:LF_1500}}
\end{figure*}


Fig.\,\ref{fig:LF_1500} shows the predicted 1500{\AA} luminosity functions at three redshift intervals. Intrinsic luminosity functions are shown in grey solid curves, while the black solid curves show luminosity functions considering AGN obscuration as described in Section.\,\ref{sec:obscuration}. Observational data from \citet{croom2009, 2013A&A...551A..29P} in SDSS g-band (4670{\AA}) were K-corrected to $z = 2$ and converted to rest-frame 1500{\AA} using the relation in \citet{griffin19}, $M_{1500}=M_g^{\prime}(z=2)+1.211$.
The predicted obscured 1500\,\AA\ luminosity function agrees with observations at $z = 0.4-0.68$, but falls below the observational constraints for $z \gtrsim 0.6$. 
At higher redshifts, observations align with dust-free predictions.

In summary, our bolometric and X-ray luminosity functions match observations up to redshift $z\approx1$. However, the model underpredicts the number of luminous AGNs at higher redshifts, possibly due to a lack of massive SMBHs or high accretion rate AGNs. 
The agreement between model predictions and observations is limited to $z\lesssim0.6$ at the optical band. Similarly, the model predicts too few bright AGNs compared to optical observational constraints at higher redshifts.

\subsubsection{Comparison with previous methods}\label{sec:compare}

\begin{figure*}
\includegraphics[width=2\columnwidth]{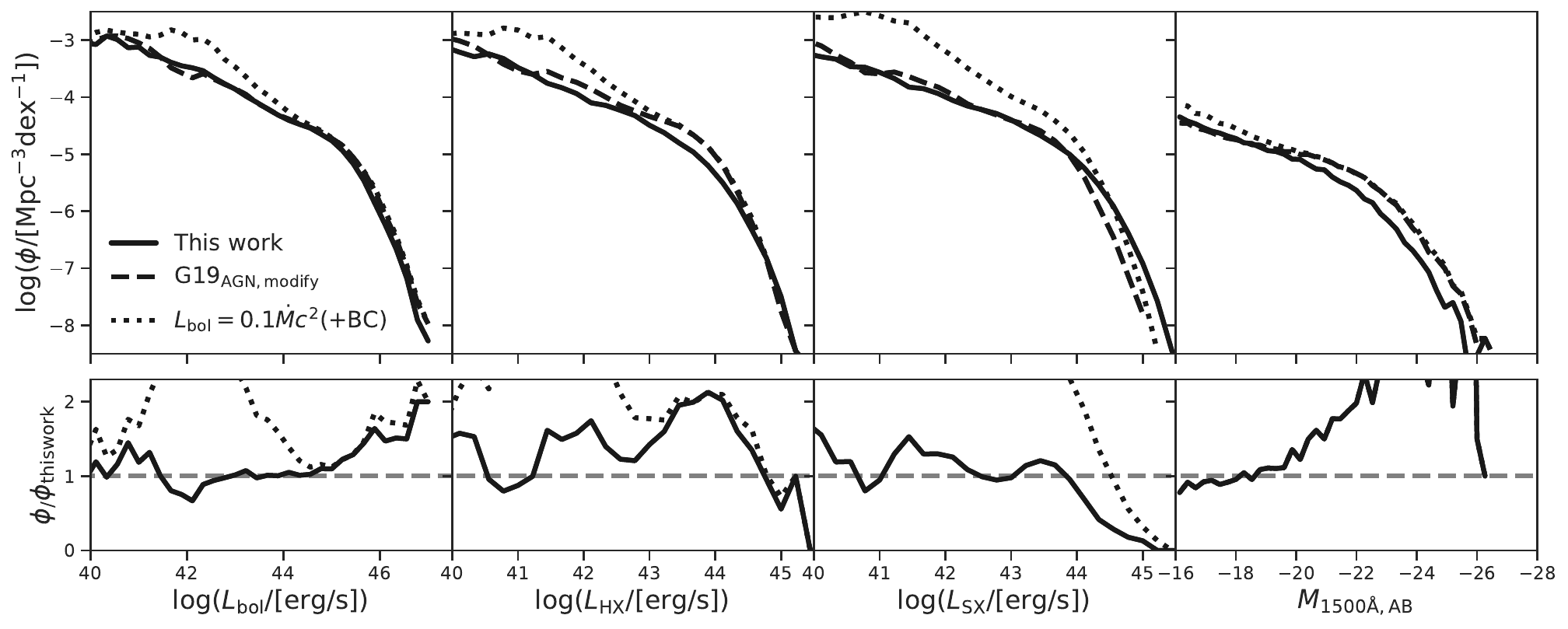}
\caption{\textbf{AGN luminosity functions with different models.} The first row displays predicted luminosity functions from our model (solid line), the $\mathrm{G19_{AGN,modify}}$ method (dashed line) with an accretion-rate-dependent $\eta$, and a model with $\eta=0.1$ (dotted line). It shows bolometric, hard X-ray, soft X-ray, and 1500\AA~ magnitude luminosity functions from left to right. In our model, photometric luminosities are directly calculated from the SED, while in the other two models, are derived from bolometric luminosities using \citet{marconi2004}. The second row illustrates the ratio of AGN luminosity functions between our model and the two mentioned models ( $\mathrm{G19_{AGN,modify}}$ model in solid black and $\eta=0.1$ model in dashed black). A horizontal grey dashed line at $\phi/\phi_{\rm thiswork}=1$ is included for reference.
\label{fig:compare}}
\end{figure*}

Previous studies often assume the bolometric luminosity is proportional to the accretion rate, defined as $L_\mathrm{bol} = \eta\dot{M}{c}^2$, where the radiative efficiency $\eta$ can vary with accretion rate and parameters in accretion models. 
A common practice is to simplify by setting $\eta=0.1$. \citet{griffin19} implemented a more sophisticated model, applying different bolometric corrections for low (ADAF), moderate (standard disk), and high (slim disk) accretion rates. 
Here we adjust the AGN model parameters ($\alpha, \beta, \delta$) in \citet{griffin19}, as well as their critical accretion rate \citep[0.01 in the original work by][]{griffin19} to be the same as those adopted in this work.
Both of these AGN models are then combined with the SMBH growth history in our semi-analytic galaxy catalog to derive the corresponding bolometric luminosities. When comparing observed luminosity functions in specific bands, many cosmological simulations use bolometric corrections to connect bolometric luminosity to photometric measurements. In practice, we combine the bolometric corrections from \citet{marconi2004} with the bolometric luminosities given by these two AGN models to predict the corresponding photometric luminosities. We refer to the former as the $\eta=0.1$ model and the latter as the $\mathrm{G19_{AGN,modify}}$ model.

In our model, bolometric and photometric luminosities are determined in the same manner by integrating over the corresponding wavelengths of the AGN SEDs.
A detailed comparison of the AGN luminosity functions between our model, $\eta = 0.1$ model, and $\mathrm{G19_{AGN,modify}}$ model at $z=0$ is presented in Fig.\,\ref{fig:compare}.

Our results are in good agreement with $\mathrm{G19_{AGN,modify}}$ for bolometric luminosities $\log(L_\mathrm{bol}/\mathrm{[erg\,s^{-1}]})\gtrsim42$, and with the $\eta=0.1$ model for luminosities $\log{(L_\mathrm{bol}/\mathrm{[erg\,s^{-1}]})}\gtrsim44$. Both $\mathrm{G19_{AGN,modify}}$ and the $\eta=0.1$ model predict higher number densities at lower luminosities compared to our model.
For the hard X-ray luminosity functions, $\mathrm{G19_{AGN,modify}}$ and $\eta=0.1$ model agree with each other at $\log(L_\mathrm{HX}/\mathrm{[erg\,s^{-1}]})\gtrsim43$, whereas our prediction only matches theirs at the very bright end, $\log(L_\mathrm{HX}/\mathrm{[erg\,s^{-1}]})\gtrsim44.5$. At lower luminosities, our predicted luminosity function is lower compared to both $\mathrm{G19_{AGN,modify}}$ and $\eta=0.1$ model predictions.
The $\eta=0.1$ model predicts higher number densities for soft X-ray luminosity functions across all luminosities compared to $\mathrm{G19_{AGN,modify}}$. Our model aligns with $\mathrm{G19_{AGN,modify}}$ in the intermediate luminosity range $\log{(L_\mathrm{SX}/\mathrm{[erg\,s^{-1}]})}=41-44$, and matches the predictions of the $\eta=0.1$ model at the bright end with log$(L_\mathrm{SX}/\mathrm{[erg\,s^{-1}]})\gtrsim44.5$.
Our model predicts lower luminosity functions in the optical band compared to the $\eta=0.1$ model and $\mathrm{G19_{AGN,modify}}$ at $M_\mathrm{1500\text{\AA}, AB}\lesssim-20$. 
At fainter magnitudes, our predictions align with the $\mathrm{G19_{AGN,modify}}$ model but are lower than the $\eta=0.1$ model. 
In general, our model agrees more closely with the $\mathrm{G19_{AGN,modify}}$ than with the $\eta=0.1$ model.


\begin{figure*}
\includegraphics[width=2\columnwidth]{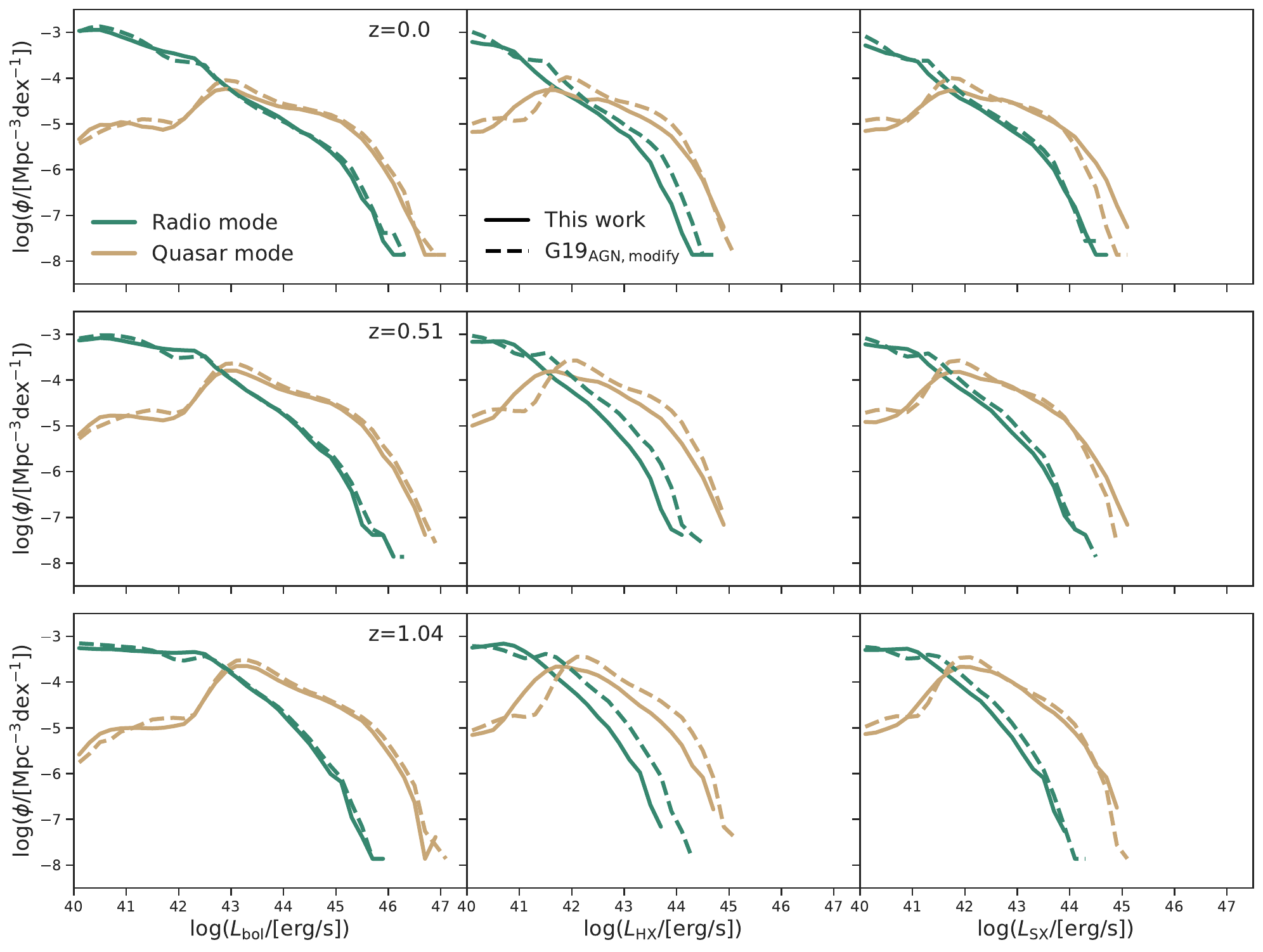}
\caption{\textbf{Detailed comparison between fuelling modes with the $\mathrm{G19_{AGN,modify}}$ model} The top, middle, and bottom rows present model predictions at redshifts $z=0, 0.51, 1.04$ respectively. From left to right, the columns represent bolometric, hard X-ray, and soft X-ray luminosity functions. Contributions from quasar-mode and radio-mode are shown as yellow and green curves respectively. Solid curves represent results from our models, while the dashed curves with the same color correspond to the $\mathrm{G19_{AGN,modify}}$ model result.
\label{fig:compareG19}}
\end{figure*}

\citet{griffin19} employed \textsc{galform} semi-analytic model of galaxy formation and high-resolution Millennium-style simulation to predict properties of AGN and SMBH. In \textsc{galform}, SMBH fueling occurs through three modes: starburst triggered by galaxy merger, starburst triggered by disk instability, and hot halo mode gas accretion. Among these, the merger-triggered starburst (stb:merger) and hot halo accretion (hh) could be interpreted as counterparts to the quasar-mode and radio-mode in \textsc{L-Galaxies} as described in Section.\,\ref{sec:SMBH_growth}. However, the disk instability-triggered starburst (stb:DI) mode is absent in the \textsc{L-Galaxies} models used in this work.
Fig.\,\ref{fig:compareG19} presents a detailed comparison of bolometric and hard/soft X-ray luminosity functions between the $\mathrm{G19_{AGN,modify}}$ model and our SED model at redshifts $z=0, 0.5, 1.0$.
Contributions of radio- and quasar-modes are shown in different colors, with solid curves representing our results, and dashed curves representing results calculated with the $\mathrm{G19_{AGN,modify}}$ model.
Generally, bolometric luminosity functions predicted by the two models align closely at all redshifts. Our model predicts a slightly lower number density for the hard X-ray luminosity functions and a slightly higher number density for the soft X-ray luminosity functions. These differences are more notable in the quasar-mode regime, and the exact value depends on the accretion state of SMBHs at certain redshifts.
Similar to those presented in \citet{griffin19} (see their Fig.\,15), the radio-mode (hh) dominates faint ends of luminosity functions, whose contribution is exceeded by the quasar-mode (stb:merg) at the bright ends. Notably, however, in \citet{griffin19}, the contribution from disk instability-triggered starburst fueling mode has a non-negligible effect on the overall luminosity function at all redshifts, and even dominates over the hot halo and merger-triggered starburst across all luminosity at higher redshift. We suggest that the absence of disk instability could potentially explain the underestimation of luminosity functions at high redshift in Section \ref{sec:LF}. Nonetheless, it is important to note that the implementation underlying the radio/hh and quasar/stb:merg modes is not entirely identical.

\section{conclusion} \label{sec:conclusion}

Reliable constraints on AGN properties depend on proper conversion of black hole accretion into observable properties, which is crucial for cosmological simulations. In this study, we utilize an advanced supermassive black hole (SMBH) accretion disk model to compute the accretion flow structure and AGN spectral energy distribution (SED) across a wide range of black hole masses and Eddington-normalized accretion rates. 
Integrating this model with the semi-analytical model \textsc{L-Galaxies} and the Millennium dark matter simulation, we study various AGN properties, including bolometric corrections, AGN radiative efficiencies, AGN luminosity functions, and AGN fractions.

Our model computes the spectral energy distribution of AGNs using SMBH masses and accretion rates across a very broad range. It combines different models for various accretion rates: Advection-Dominated Accretion Flow with a thin disk for low rates, and a modified magnetic reconnection-heated disk-corona model for higher rates. This SED model is more sophisticated and self-consistent than those previously used in most cosmological simulations and galaxy formation models, which typically rely on relatively simple bolometric corrections between the SMBH accretion rate and bolometric luminosity, followed by simple conversions of the bolometric luminosity to photometric luminosities in specific bands.

We found that the radiative efficiency depends strongly on the accretion rate but weakly on the SMBH mass. SMBH mass and Eddington-normalized accretion rate play comparable roles in determining the bolometric luminosity. Our model aligns with simpler treatments in the literature for intermediate accretion rates ($-2<\log \dot m<0$). However, at higher accretion rates, most models predict higher radiation efficiencies by neglecting the photon-trapping effect; while at lower accretion rates, compared to our model, some models show higher efficiencies and others lower. However, a more complex model in \citet{griffin19} matches ours across all accretion rates.

We established new bolometric corrections linking bolometric luminosities with X-rays and optical luminosities, considering the intrinsic scatter of the SED for a given bolometric luminosity. Our model predicts lower X-ray luminosities at low bolometric luminosities in comparison with SED models widely used in the literature.

AGN luminosity comparison between semi-analytical catalogs and observations is summarized as follows.
\begin{enumerate}
\item We reproduced the active AGN fraction in massive galaxies over the redshift range $z = 0-2.5$.

\item We replicated the bolometric and hard/soft X-ray luminosity functions up to $z = 1$. Beyond this redshift, the model predicts fewer X-ray sources compared to observations at very bright ends.
\item {The predicted obscured optical luminosity functions agree well with the observations at $z \lesssim 0.6$. But the model predicts fewer bright optical sources at higher redshift, similar to that of the bolometric and X-ray bands.
}
\item 
ADAF+disk/radio-mode dominates AGN luminosity function at the faint end, i.e., below $\log(L_\mathrm{bol}/\mathrm{[erg\,s^{-1}]})\approx42$; While the disk-corona/quasar-mode dominates the bright end. The disk-corona mode starts to dominate at even fainter AGNs compared to the quasar-mode, indicating a high accretion rate triggered by the radio-mode accretion. The influence of redshift on the contributions from various accretion modes is noticeable in the intermediate luminosity range between $\log(L_\mathrm{bol}/\mathrm{[erg\,s^{-1}]})\approx41.5-45$.
\end{enumerate}


In general, the AGN luminosity function matches observations below $z \approx 1$ but appears lower at higher redshifts. This discrepancy at high redshifts could arise from inadequate treatments of SMBH growth or observational bias.
As mentioned earlier, this work does not account for SMBH spin, which could influence our model predictions, particularly at high redshift, due to the dependence of radiative efficiency on black hole spin. \citet{2012ApJ...749..187L} suggested that SMBHs experience a period of spin-up through prolonged accretion, and a period of spin-down due to random, episodic accretion towards low redshift, indicating that the radiative efficiency at high redshift could be systematically higher than that of the zero-spin scenario considered in this work. We suggest that the absence of SMBH spin in the \citetalias{Henriques15} \textsc{L-Galaxies} semi-analytic model could partially explain the deviation between model predictions and observational data. In addition, SMBH growth driven by disk instabilities could also play a significant role in SMBH growth at high redshifts, a factor that is not considered in this work.
The upcoming data from X-ray  \citep{Lynx2018}, optical and infrared bands (EUCLID \citep{Euclid2019}, CSST \citep{2022MNRAS.511.1830C}, WFIRST \citep{2019arXiv190205569A}, gravitational waves (LIGO \citep{Abbott2016}, Laser Interferometer Space Antenna (LISA) \citep{amaro2017}, PTA \citep{2023A&A...678A..48E,2023A&A...678A..49E,2023A&A...678A..50E,2024A&A...685A..94E} ) would provide more constraints on the formation models, accretion mode and the luminosities of the SMBH.

\section*{Data Availability}

The interpolated functions for calculating AGN SED are available in the GitHub repository\footnote{\url{https://github.com/SuTong1999/agnSED}}.
Additional data produced in this study are available upon reasonable request to the corresponding author.

\section*{Acknowledgement}
We thank Tao Wang, Ye Feng, and Teng Liu for helpful discussions and comments. 
This work is supported by the National Natural Science Foundation of China (NSFC) (Nos. 12425303, 12033008, 12588202), the National SKA Program of China (No. 2022SKA0110201),
the CAS Project for Young Scientists in Basic Research grant No. YSBR-062, the K.C.Wong Education Foundation, and the Strategic Priority Research Program of the Chinese Academy of Sciences, Grant No.XDB0500203.
QG acknowledges the hospitality of the International Centre of Supernovae (ICESUN), Yunnan Key Laboratory at Yunnan Observatories, Chinese Academy of Sciences, and the European Union's HORIZON-MSCA-2021-SE-01 Research and Innovation programme under the Marie Sklodowska-Curie grant agreement number 101086388. 
EQ was supported by the Strategic Priority Research Program of the Chinese Academy of Science (grant No. XDB0550200), the National Key R\&D Program of China (No. 2023YFA1607903), and the National Natural Science Foundation of China (grant No. 12333004).
LCH was supported by the National Science Foundation of China (12233001) and the China Manned Space Program (CMS-CSST-2025-A09).
CGL acknowledges support from STFC grant ST/X001075/1.

\appendix

{

\section{SED illuminations}\label{sec:SED4panel}

\begin{figure*}
\includegraphics[width=2\columnwidth]{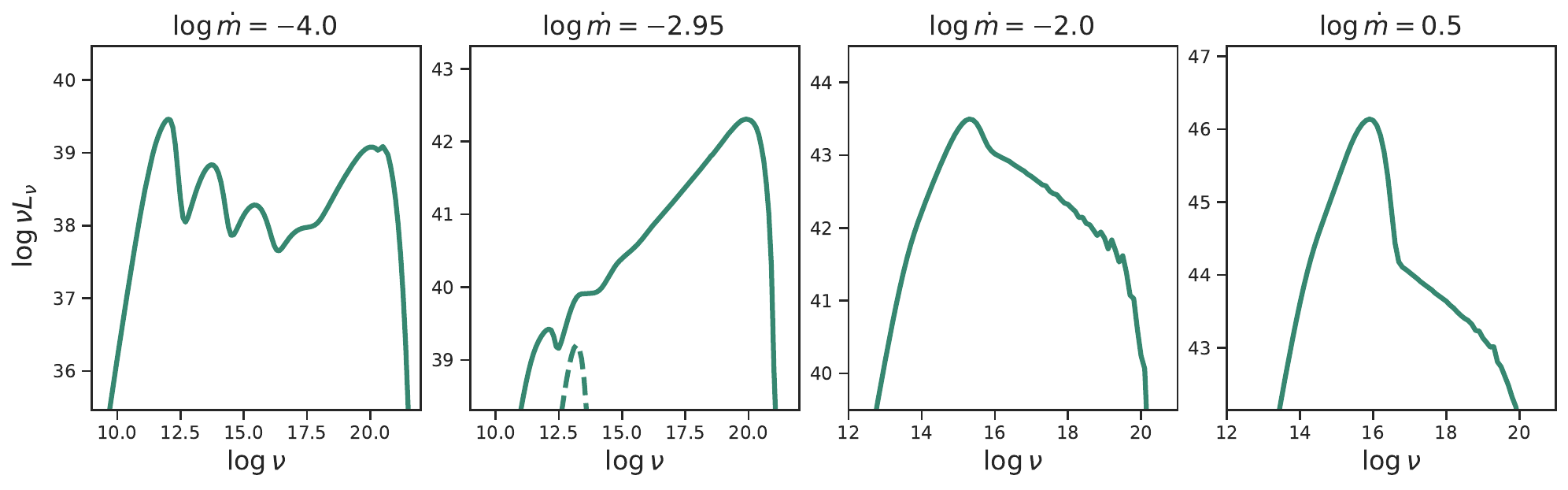}
\caption{\textbf{SED evolution with increasing Eddington-normalized accretion rate} Example SEDs of SMBHs with $\log(M_\mathrm{SMBH}/\mathrm{M_\odot})=8$ and Eddington-normalized accretion rates of $\log\dot{m}=-4.0,\, -2.95,\, -2.0,\, 0.5$, shown from left to right, corresponding to the four accretion regimes in Fig.\,\ref{fig:geometry}.
\label{fig:SED4panel}}
\end{figure*}

Fig.\,\ref{fig:SED4panel} shows the SEDs of a $10^8\,\mathrm{M_\odot}$ SMBH with Eddington-normalized accretion rates of $\log\dot{m}=-4.0,\, -2.95,\, -2.0,\, 0.5$, from the left panel to the right respectively. Each panel corresponds to the state shown in Fig.\,\ref{fig:geometry}: pure ADAF; ADAF+disk; disk-corona system; disk-corona with inner slim component.
}.

\section{AGN SED Variations} \label{sec:bols}
\begin{figure}
\includegraphics[width=\columnwidth]{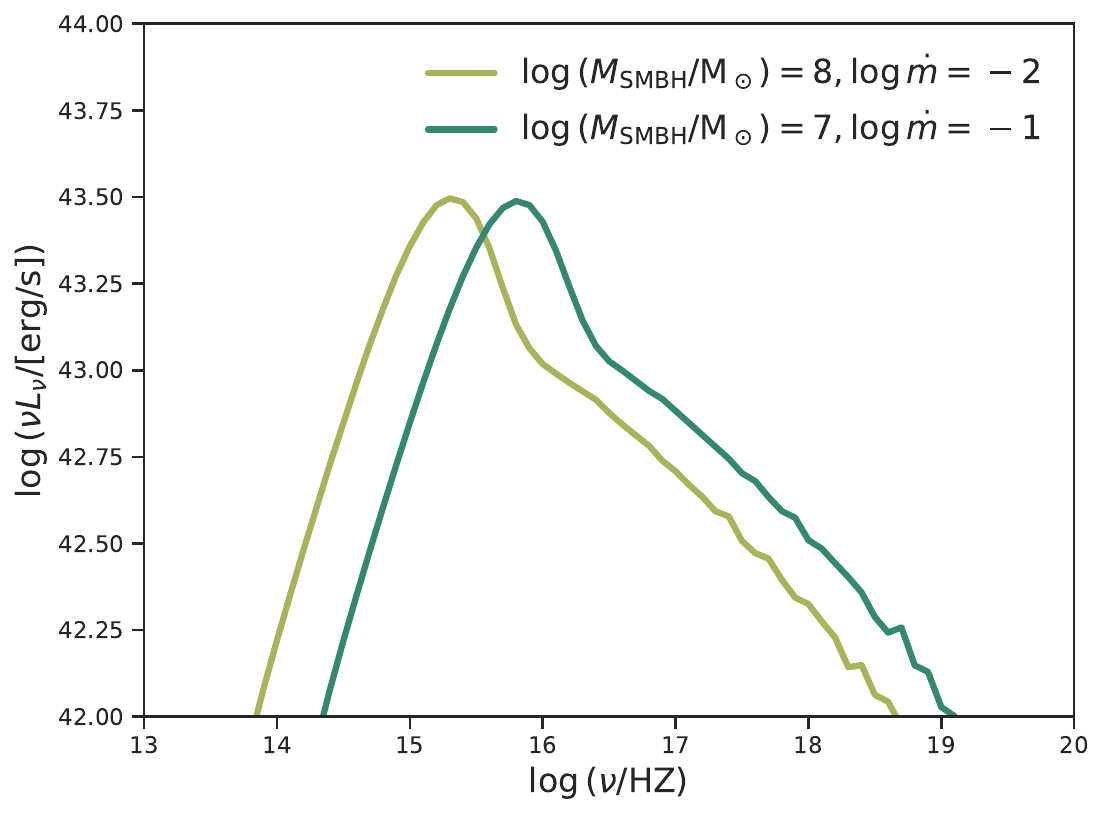}
\caption{\textbf{An example of two SMBHs with different masses and Eddington-normalized accretion rates but with identical integrated bolometric luminosities.}
SEDs of a $M_\mathrm{SMBH}=10^8\,\mathrm{M_\odot}, \dot{m}=10^{-2}$ SMBH and a $M_\mathrm{SMBH}=10^7\,\mathrm{M_\odot}, \dot{m}=10^{-1}$ SMBH respectively.
The corresponding bolometric luminosities are roughly the same with $L_\mathrm{bol}\approx10^{43.9}\mathrm{erg\,s^{-1}}$, while the luminosities in specific bands have non-negligible differences.
\label{fig:same_bol}}
\end{figure}

Higher SMBH mass and accretion rate lead to increased bolometric luminosity. However, the spectrum peak frequency increases with accretion rate but decreases with SMBH mass. 
This could be interpreted as follows: provided the SMBHs have the same Schwarzschild radius-normalized disk size, the mass accretion rate per unit area scales as $\propto \dot{M}_\mathrm{acc}/R_\mathrm{S}^2\propto \dot{m}/M_\mathrm{SMBH}$. Consequently, given an Eddington-normalized accretion rate, a larger SMBH tends to have a lower accretion rate per unit area, leading to a cooler temperature and causing the peak frequency to shift to lower values. Meanwhile, the bolometric luminosity scales as $\propto\dot M\propto M_\mathrm{SMBH}\dot{m}$ (for the disk-corona model). Therefore, a larger SMBH mass and a smaller Eddington-normalized accretion rate can produce the same bolometric luminosity but with a lower peak frequency.
Fig.\,\ref{fig:same_bol} shows two distinct SEDs for black holes of different masses and accretion rates but with the same bolometric luminosity of approximately $\log{(L_\mathrm{bol}/\mathrm{[erg\,s^{-1}]})} \approx 43.9$. The peak frequencies of the SEDs are notably distinct, leading to significant differences in their photometric luminosities in different bands.

\section{Influence of model parameters on the ADAF model}\label{append:ADAF_para}

\begin{figure*}
\includegraphics[width=2\columnwidth]{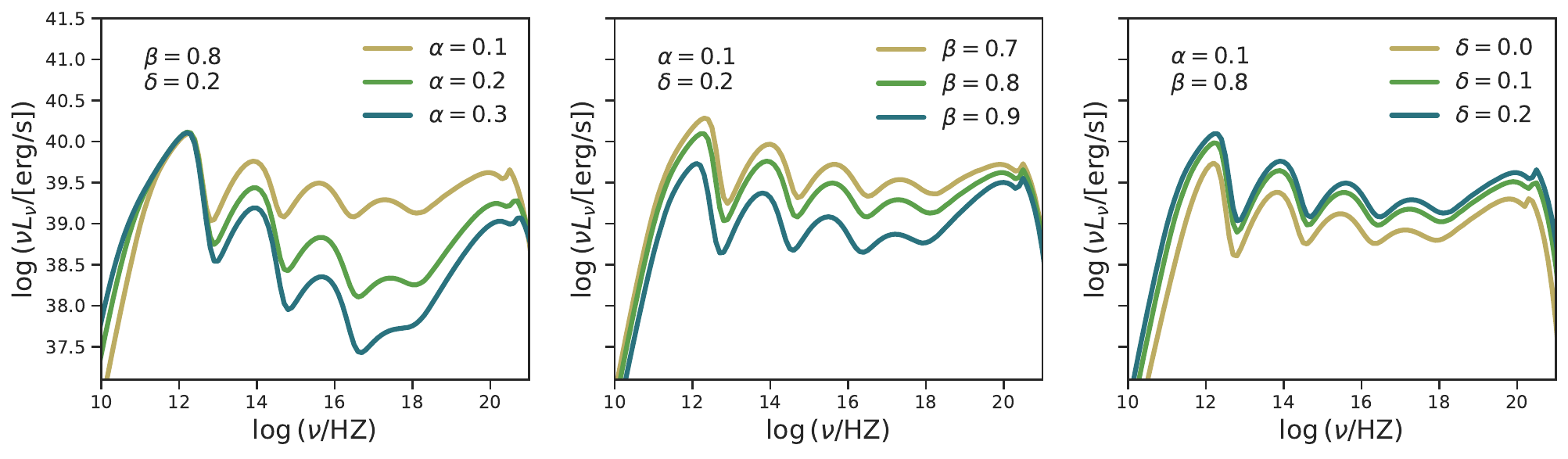}
\caption{\textbf{Influence of model parameters on the ADAF model.}
SEDs of a $\log (M_\mathrm{SMBH}/\mathrm{M_\odot})=8, \log{\dot{m}}=-3.5$ SMBH by varying each one of the three model parameters in each panel. 
\textit{Left:} SEDs of the ADAF model by varying the viscosity parameter $\alpha=0.1, 0.2, 0.3$, and the other two parameters are set to be $\beta=0.8, \delta=0.2$; 
\textit{Middle}: SEDs of the ADAF model by varying the magnetic parameter $\beta=0.7, 0.8, 0.9$, and the other two parameters are set to be $\alpha=0.1, \delta=0.2$; \textit{Right}: SEDs of the ADAF model by varying the $\delta$ parameter $\delta=0.0, 0.1, 0.2$, and the other two parameters are set to be $\alpha=0.1, \beta=0.8$
\label{fig:ADAF_para}}
\end{figure*}

As described in the main text, in addition to the input variables SMBH mass and accretion rate, three model parameters govern the structure of the ADAF model - the viscosity parameter $\alpha$, magnetic parameter $\beta$, and the dissipation parameter $\delta$.
Fig.\,\ref{fig:ADAF_para} shows the different SEDs of an SMBH with fixed mass and Eddington-normalized accretion rate $\log{(M_\mathrm{SMBH}/\mathrm{M_\odot})}=8.0,\ \log{\dot{m}}=-3.5$. Each model parameter is varied within reasonable ranges from the left to the right, whilst the other two are fixed at their fiducial values ($\alpha=0.1, \beta=0.8, \delta=0.2$). 
As shown in the figure, the luminosity of the ADAF increases with decreasing $\alpha$ and $\beta$, which can be understood as $\alpha$ being anti-correlated with electron density $\rho$ \citep[see the self-similar solution][]{narayan&yi1995a,narayan&yi1995b}, and from the definition of $\beta$, smaller value means stronger magnetic field, which is directly related to the intensity of Synchrotron emission; and because $\delta$ adds heating term to the electrons in the ADAF, the luminosity has to increase to reach energy equilibrium.

\begin{figure}
\includegraphics[width=\columnwidth]{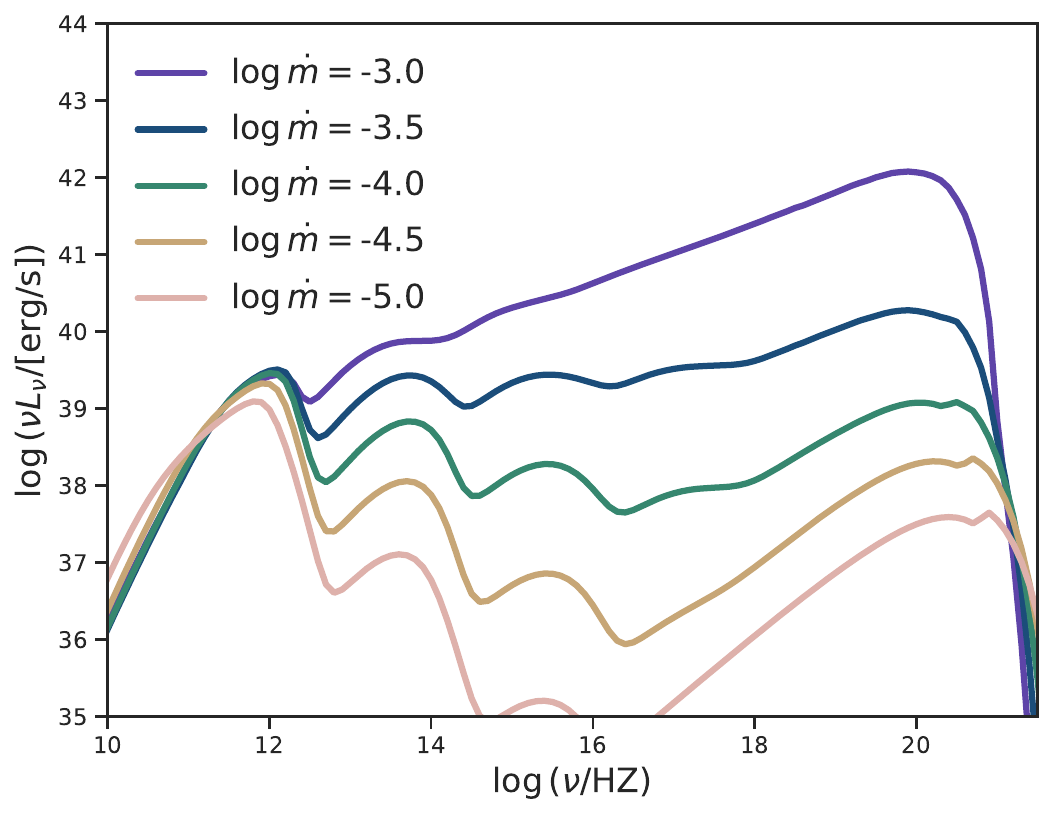}
\caption{\textbf{SEDs of the ADAF model of a $10^{8}\,\mathrm{M}_\odot$ SMBH.} The SEDs are calculated using the ADAF model for accretion rates ranging from $\log{\dot{m}} = -3.0$ to $-5.0$, denoted with solid curves
\label{fig:SED_adaf}}
\end{figure}

In this work, we did not modify the ADAF model; its SED remains consistent with those presented in previous literature. \citep[e.g.,][]{narayan&yi1995a}. For clarity, we show SEDs of a $10^8\mathrm{M_\odot}$ SMBH, with Eddington-normalized accretion rate $\log{\dot m}=-3,\,-3.5,\,-4,\,-4.5,\,-5$ in Fig.\,\ref{fig:SED_adaf}.

\section{The disk-corona model}\label{sec:append_diskcor}
\subsection{Self-similar solution of the slim disk}\label{sec:appendSlim}

Since the concept of slim disk has been proposed by \citet[][]{abrawicz1988}, great efforts have been made to solve its structure and emergent spectrum \citep[e.g.,][]{2000PASJ...52..499M, watarai2000, 2001ApJ...549L..77W}.
Detailed calculation for the structure and the emergent spectrum of the slim disk is very complicated, depending on how to treat the radiative transfer, the effect of wind, etc, which has exceeded the scope of this paper. 
We take the modified self-similar solution as a zeroth-order approximation, which captures the basic properties of the slim disk, such as the photon-trapping effect and the saturated luminosity, and is sufficient for a statistical study of the accretion flow in AGN. 
We adopt most of the assumptions and procedures outlined in \citet{watarai2006}, with two modifications as described in the main text. For clarity, the modified formulae are outlined in this section.
In the slim disk scenario, 
the radiative flux at the disk surface can be estimated using Rosseland approximation \citep{1977PThPh..58.1191H}, 
\begin{equation}\label{eq:slimT}
\begin{split}
F(H) &= -\frac{16}{3} \frac{\sigma T^3(z)}{\kappa_\mathrm{es} \rho(z)} \frac{\partial T}{\partial z} \bigg|_{z=H} 
     = \sigma T^4_\mathrm{eff} = \frac{1}{2}Q_\mathrm{rad}.
\end{split}
\end{equation}
Where $H$ is the half thickness of the disk, $\Pi$, $\Sigma$ are the vertical integrated pressure and density (refer to \citet{watarai2006} for details), $\sigma$ is the Stephan-Boltzmann constant, $a=\frac{4}{c}\sigma$ is the radiation constant, $\kappa_\mathrm{es}$ is the electron scattering opacity, and $I_3=16/35, \,I_4=128/315$ are two numerical factors \citep{2008bhad.book.....K}.
Note that in \citet{watarai2006}, the $\partial T(z)/\partial z$ term is approximated to be $T_0/H$, and the $\rho_0 H$ term in the denominator is approximated to be $\Sigma/2$. While from the vertical profiles of $T,\,\rho$ assuming polytropic relation, $\partial T(z)/\partial z=2z/H^2$ and $\rho_0 H=\Sigma/2I_3$, which introduces $2I_3$ difference between Eq.\,\ref{eq:slimT} and Eq.\,19 in \citet{watarai2006}.

The effective temperature $T_\mathrm{eff}$ 
can be calculated as,
\begin{equation}\label{eq:append_slimT}
\begin{aligned}
T_{\mathrm{eff}} = \left[\frac{32 I_3}{\kappa_{\mathrm{es}} I_4 a} \sqrt{\frac{\mathcal{B} \Gamma_{\Omega}}{(2 N+3) \xi}}\right]^{1 / 4} f^{1 / 8} R^{1 / 4} (\Omega\,\Omega_{\mathrm{K}})^{1 / 4}
\end{aligned}
\end{equation}
in which $N=3$, $\xi=1.5$ are dimensionless quantities.
We follow \citet{watarai2006} to assume the angular velocity to be $\Omega=\Omega_0\,\Omega_\mathrm{K}$, where $\Omega_0=1$, and $\Omega_\mathrm{K}$ is the Keplerian angular velocity in Newtonian potential $\Omega_\mathrm{K}=\sqrt{GM/R^3}$, calculated by the gravitational constant $G$, SMBH mass $M$ and radial distance $R=R_\mathrm{S}r$. With the definition of $\Omega$, the boundary term $\mathcal{B}=1-l_\mathrm{in}/l$ can be calculated as $\mathcal{B}=1-\sqrt{R_\mathrm{in}/R}=1-\sqrt{3/r}$, where $l=R^2\Omega$ is the angular momentum; and $\Gamma_\Omega=-d\ln{\Omega}/d\ln{r}=1.5$. $f$ is the fraction of viscous heating that is advection-cooled, the $2I_3$ factor difference in Eq.\,\ref{eq:append_slimT} will also result in an order-of-unity difference in the value of $f$ compared to \citet{watarai2006}.

\subsection{Influence of model parameters on the disk-corona model}\label{sec:diskcor_para}

\begin{figure*}
\includegraphics[width=2\columnwidth]{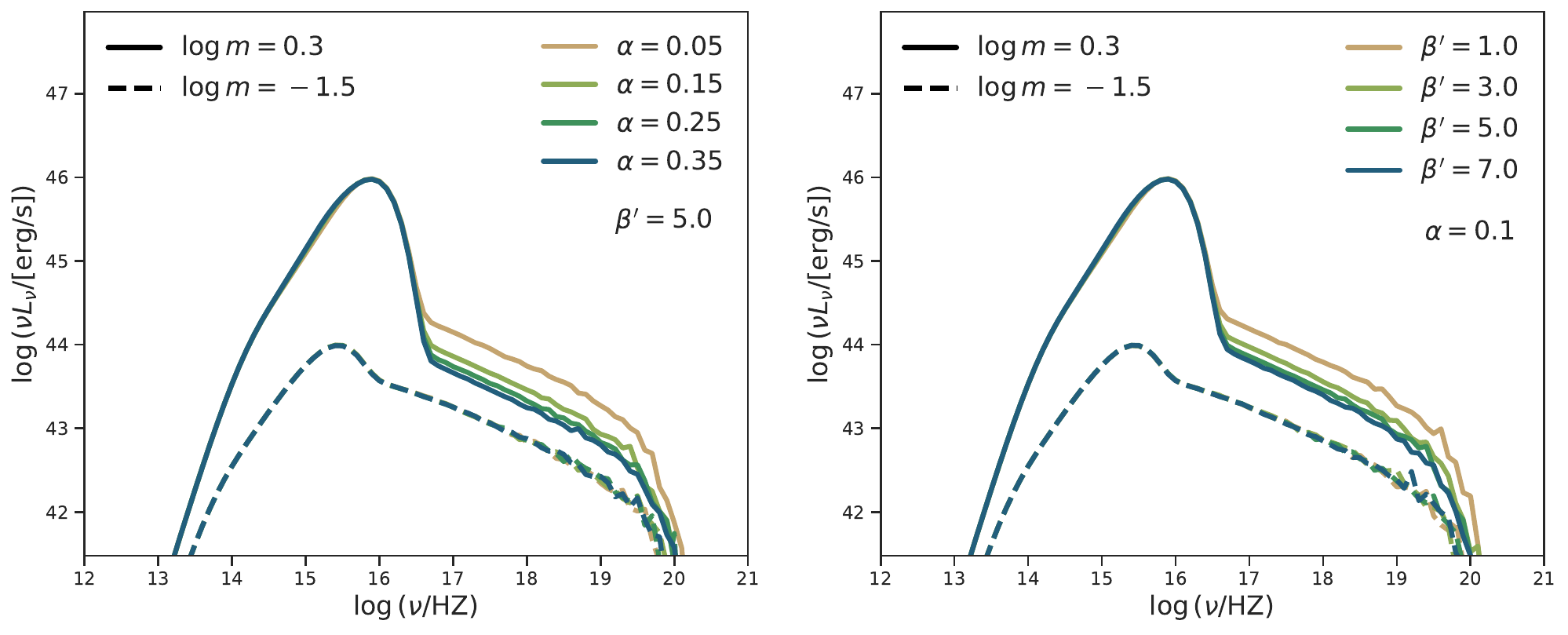}
\caption{\textbf{Influence of model parameters on the disk-corona SED.} 
The SMBH mass is taken to be $10^8\,\mathrm{M_\odot}$, and SMBH accretion rates are taken to be $\log{\dot{m}}=-1.5, 0.3$. The solid curve set represents the $\log{m}=-1.5$ case, and the dashed curve set represents the $\log{m}=0.3$ case.
\textit{Left:} SEDs of the disk-corona model by varying the viscosity parameter $\alpha=0.05, 0.15, 0.25, 0.35$ with fixed $\beta'=5.0$; 
\textit{Right}: SEDs of the disk-corona model by varying the magnetic parameter $\beta'=2.0, 4.0, 6.0, 8.0$ with fixed $\alpha=0.1$.
\label{fig:diskcor_para}}
\end{figure*}

In addition to the input SMBH mass and accretion rate, two model parameters govern the structure of the disk-corona model, i.e., the viscosity parameter $\alpha$, and the magnetic parameter $\beta^\prime$.
Fig.\,\ref{fig:diskcor_para} shows the influence of model parameters $\alpha$ and $\beta^\prime$ on the disk-corona SED, where the SMBH mass is taken to be $10^8\,\mathrm{M_\odot}$ and SMBH accretion rates to be $\log{\dot{m}}=-1.5, 0.3$, denoted with dashed and solid curve sets respectively. Each model parameter is varied within its reasonable ranges while keeping the other fixed ($\beta'=5.0$ for the left panel and $\alpha=0.1, $ for the right panel).
For the $\log{\dot{m}}=0.0$ case, the strength of corona (or hardness of the spectrum) is anti-correlated with both $\alpha$ and $\beta$ for similar reasons as in the ADAF scenario, note that because the high-energy emission contribution is almost negligible to the total luminosity, varying model parameters do not alter the bolometric luminosity meaningfully. However, in the $\log{\dot{m}}=-1.5$ case, the emergent spectra are insensitive to the value of $\alpha$ and $\beta$.

\section{SED of individual source}\label{append:sed_example}

\begin{figure*} 
\includegraphics[width=2\columnwidth]{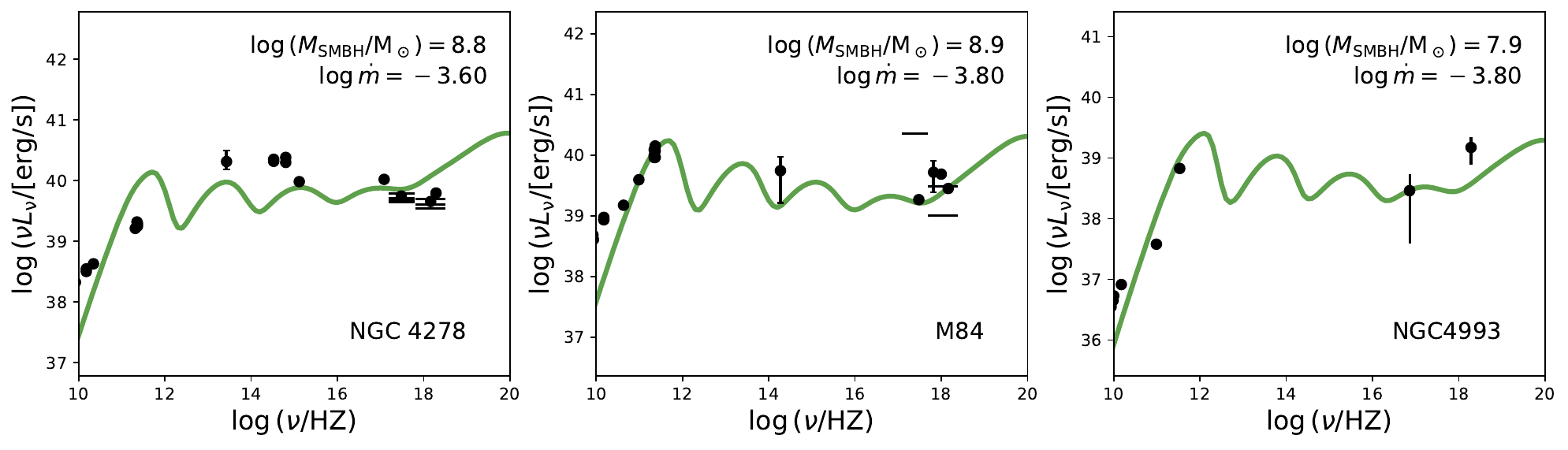}
\caption{\textbf{SEDs of NGC4278, M84 and NGC4993.} 
The solid curves show three SEDs that are broadly consistent with observational data. Model parameters are taken as values adopted in this work  ($\alpha=0.05,\,\beta=0.95,\,\delta=0.2$). The SMBH mass and Eddington-normalized accretion rate for each case are shown in their respective panels.
\textit{Left:} NGC4278, observations are collected from \citet{Bandyopadhyay2019, Chen2023, Hernandez-Garcia2013}. \textit{Middle:} M84, observations are collected from \citet{Bandyopadhyay2019, Bambic2023, Bogdan2011}. \textit{Right:} NGC4993, observations are collected from \citet{Wu2018}.
\label{fig:source_adaf}}
\end{figure*}
\begin{figure*}
\includegraphics[width=2\columnwidth]{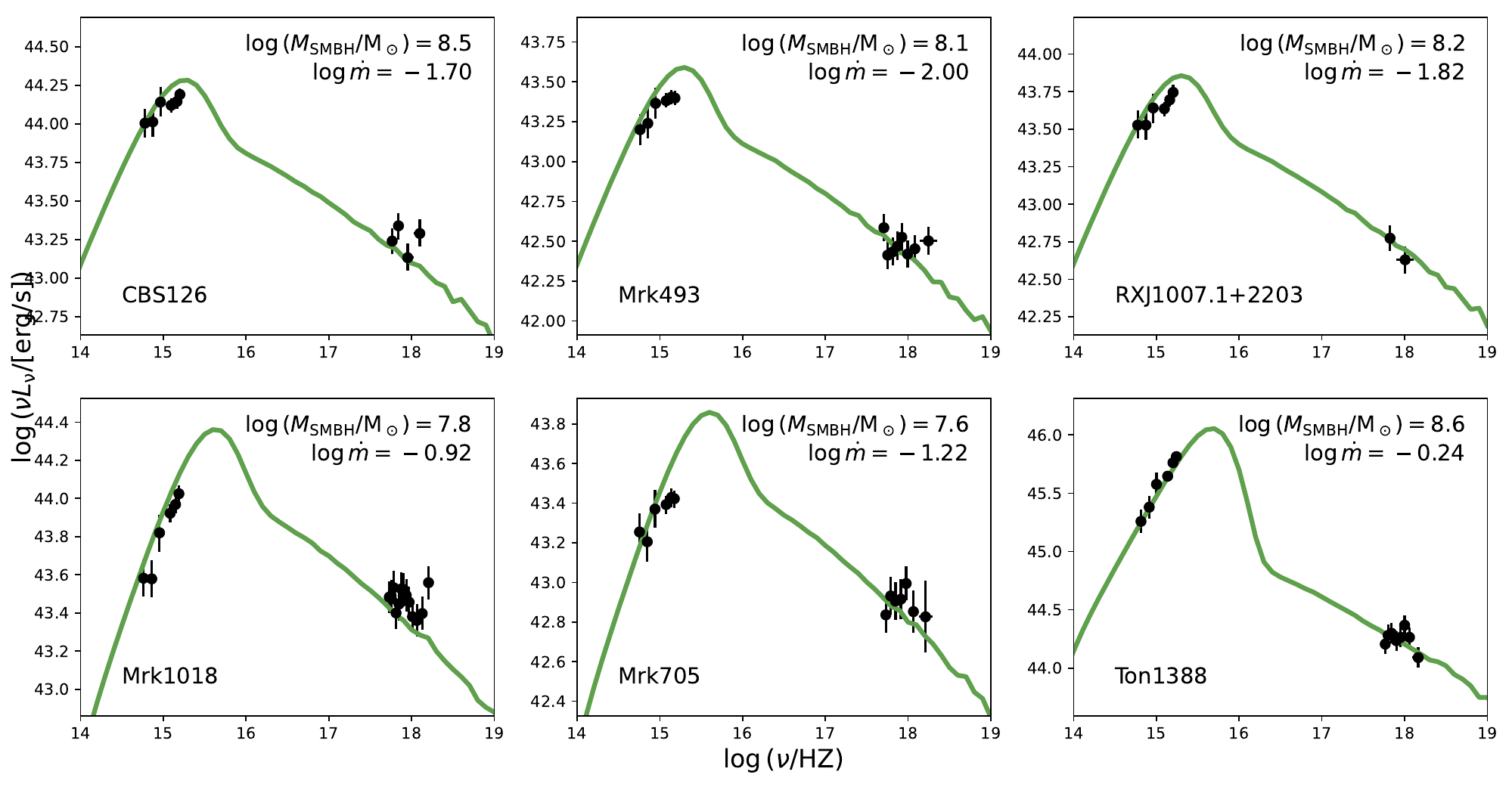}
\caption{ \textbf{SEDs of CBS126, Mrk493, RXJ1007.1+2203, Mrk1018, Mrk705, Ton1388.} Model parameters are taken as values adopted in this work ($\alpha=0.05,\,\beta^\prime=8$). The SMBH mass and Eddington-normalized accretion rate for each case are shown in their respective panels. Observations are collected from \citet{Cheng2020, Cheng2019}.
\label{fig:source_diskcor}}
\end{figure*}

Fig.\,\ref{fig:source_adaf} displays the SEDs of three low luminosity AGNs – NGC4278, M84, and NGC4993 from left to right. Model parameters are taken as values adopted in this work  ($\alpha=0.05,\,\beta=0.95,\,\delta=0.2$). The SMBH mass and Eddington-normalized accretion rate for each case are shown in their respective panels, 
which are adjusted based on the best-fitted values from \citet{Bandyopadhyay2019, Wu2018}, who employ similar, though not identical SED models and perform detailed spectral fitting for individual sources.
Observations are collected from \citet{Bandyopadhyay2019, Chen2023, Hernandez-Garcia2013, Bambic2023, Bogdan2011, Wu2018}.
Similar comparisons are conducted for more luminous AGNs, as shown in Fig.\,\ref{fig:source_diskcor}, based on the disk-corona model, with the SMBH mass and accretion rate shown in the upper-right corner of each panel, and model parameters are taken as values adopted in this work  ($\alpha=0.05,\beta^\prime=0.95$).
Similarly, the SMBH masses and accretion rates are adjusted based on the best-fitted results from \citet{Cheng2020}, who employ similar, though not identical SED models and perform detailed spectral fitting for individual sources.
Observational data for CBS126, Mrk493, RXJ1007.1+2203, Mrk1018, Mrk705, and Ton1388 are taken from \citet{Cheng2020, Cheng2019}. In general, with fixed parameters $\alpha$, $\beta$, $\beta^\prime$, and $\delta$ as adopted in the main text, our SED model can reproduce SEDs that broadly resemble the observed AGN SEDs across different accretion regimes, capturing the general characteristics of observed AGN SEDs. While not intended for precise fits, these results demonstrate that the model can reasonably reproduce realistic SED shapes across different accretion regimes. Table.\,\ref{tab:append_sources} summarizes the redshifts and model-predicted luminosities of each source.

\begin{table}[h!]
\centering
\begin{tabular}{cccc}
\hline
\textbf{} & \textbf{Source} & \textbf{Redshift} & \textbf{$\log (L_\mathrm{bol, model})$} \\ \hline
\multirow{6}{*}{HLAGN} & CBS126 & 0.08 & 44.8 \\ \cline{2-4}
                       & Mrk 493 & 0.03 & 44.1 \\ \cline{2-4}
                       & RXJ1007.1+2203 & 0.08 & 44.4 \\ \cline{2-4}
                       & Mrk 1018 & 0.04 & 44.9 \\ \cline{2-4}
                       & Mrk 705 & 0.03 & 44.4 \\ \cline{2-4}
                       & Ton 1388 & 0.18 & 46.4 \\ \hline
\multirow{3}{*}{LLAGN} & NGC 4278 & 0.002 & 41.5 \\ \cline{2-4}
                       & M84 & 0.005 & 41.2 \\ \cline{2-4}
                       & NGC 4993 & 0.01 & 40.3 \\ \hline

\end{tabular}
\caption{
\textbf{Redshifts and model-predicted luminosities of AGNs in Appendix.\,\ref{append:sed_example}.} Redshift estimations are taken from \citet{2012MNRAS.425.1299C, 2013ApJS..209....1F, 2016A&A...593L...8M, 2010ApJS..187...64G, 2009A&A...506.1107G, 2017ApJ...848L..28L}.}
\label{tab:append_sources}
\end{table}




\bibliography{example}{}
\bibliographystyle{aasjournal}



\end{document}